\documentclass[twocolumn]{article}
\pdfoutput=1
\usepackage{graphicx}
\usepackage{amssymb}

\begin{document}

\makeatletter
\def\imod#1{\allowbreak\mkern10mu({\operator@font mod}\,\,#1)}
\makeatother

\newcommand{\bra}[1]{\mbox{$\left\langle #1 \right|$}}
\newcommand{\ket}[1]{\mbox{$\left| #1 \right\rangle$}}
\newcommand{\braket}[2]{\mbox{$\left\langle #1 | #2 \right\rangle$}}
\newcommand{\av}[1]{\mbox{$\left| #1 \right|$}}
\newcommand{\ve}{\varepsilon}
\newcommand{\osc}{{\mbox{\rm \scriptsize osc}}}
\newcommand{\tot}{{\mbox{\rm \scriptsize tot}}}
\newcommand{\lga}{{\mbox{\rm \scriptsize LGA}}}
\newcommand{\swap}{{\mbox{\rm \scriptsize swap}}}
\newcommand{\ground}{{\mbox{\rm \scriptsize ground}}}
\newcommand{\cycle}{{\mbox{\rm \scriptsize cycle}}}
\newcommand{\particle}{{\mbox{\rm \scriptsize particle}}}
\newcommand{\internal}{{\mbox{\rm \scriptsize internal}}}
\newcommand{\nonrel}{{\mbox{\rm \scriptsize non-rel}}}
\newcommand{\twoparticles}{{\mbox{\rm \scriptsize 2p}}}
\newcommand{\block}{{\mbox{\rm \scriptsize block}}}
\newcommand{\blockchange}{{\mbox{\rm \scriptsize block-change}}}
\newcommand{\statechange}{{\mbox{\rm \scriptsize state-change}}}
\newcommand{\even}{{\mbox{\rm \scriptsize even}}}
\newcommand{\odd}{{\mbox{\rm \scriptsize odd}}}
\newcommand{\rt}{{\mbox{\rm \scriptsize R}}}
\newcommand{\lt}{{\mbox{\rm \scriptsize L}}}
\newcommand{\shift}{{\mbox{\rm \scriptsize shift}}}
\newcommand{\D}[2]{\frac{\partial #2}{\partial #1}}
\newcommand{\pp}{{\mbox{\tt \scriptsize /}}}
\newcommand{\mm}{{\mbox{\tt \scriptsize \backslash}}}

\newcommand{\sinc}{{\mbox{\rm sinc}}\:}
\newcommand{\sincs}{{\mbox{\rm sinc$^2$}}}

\renewcommand{\max}{{\mbox{\rm \scriptsize max}}}
\renewcommand{\min}{{\mbox{\rm \scriptsize min}}}

 \newcommand{\myfig}[3]{
 \begin{figure*}
  $$#2$$\relax
  \caption{#3}
 \label{fig.#1}
 \end{figure*}                 }

 \newcommand{\myfign}[3]{
 \begin{figure}
  $$#2$$\relax
  \caption{#3}
 \label{fig.#1}
 \end{figure}                 }

\setlength{\fboxsep}{.1pt}
\setlength{\fboxrule}{.1pt}

 \title{Mechanical Systems that are both Classical and Quantum}

 \author{Norman Margolus\footnote{
MIT, Cambridge MA 02139; nhm@mit.edu}}

\date{}



\maketitle

 \begin{abstract} 


{\em Quantum dynamics can be regarded as a generalization of classical
  finite-state dynamics.  This is a familiar viewpoint for workers in
  quantum computation, which encompasses classical computation as a
  special case.  Here this viewpoint is extended to mechanics, where
  classical dynamics has traditionally been viewed as a macroscopic
  approximation of quantum behavior, not as a special case.

When a classical dynamics is recast as a special case of quantum
dynamics, the quantum description can be interpreted classically.  For
example, sometimes extra information is added to the classical state
in order to construct the quantum description.  This extra information
is then eliminated by representing it in a superposition as if it were
unknown information about a classical statistical ensemble.  This
usage of superposition leads to the appearance of Fermions in the
quantum description of classical lattice-gas dynamics and turns
continuous-space descriptions of finite-state systems into
illustrations of classical sampling theory.

A direct mapping of classical systems onto quantum systems also allows
us to determine the minimum possible energy scale for a classical
dynamics, based on a localized rate of state change.  We use a
partitioning description of dynamics to define locality, and discuss
the ideal energy of two model systems.}

\end{abstract}


\section{Introduction}\label{sec.intro}

A digital photograph looks continuous but in fact, if you examine it
closely enough, you discover that there is only a finite amount of
detail.  Similarly, a digital movie looks like it is changing
continuously in time, but in fact it is actually a discrete sequence
of digital images.

Something similar is true of nature.  Although the world looks to our
senses as if it has an infinite amount of resolution in both space and
time, in fact a finite-sized physical system with a finite energy has
only a finite amount of distinguishable detail and this detail changes
at only a finite rate~\cite{max-speed}.

\subsection{A bit of history}

The finite character of the states of physical systems came as a great
surprise when it became apparent at the start of the twentieth
century.  The revolution was started by Max Planck in 1900 when he
found that he had to introduce a new constant into physics in order to
understand the thermodynamics of electromagnetic radiation in a
cavity.  The new constant fixed the statistical mechanical analysis,
but it did so by making the count of distinct possible states finite.

Planck's constant has a particularly simple interpretation in
classical statistical mechanics.  There, the number of microscopic
ways a system can realize a macroscopic state is taken to be
proportional to the volume in {\em phase space} (i.e.,
position/momentum space) of states consistent with the parameters that
define the macroscopic state.  In units where Planck's constant is
one, the phase space volume becomes the actual finite count of
distinct states.  Planck's constant is the fundamental grain size in
phase space, representing one state.

Although Planck's constant was initially introduced to fix statistical
mechanics, a new dynamical theory also grew from this beginning.  The
resulting quantum mechanics established new rules for describing the
microscopic dynamics of physical systems.  The QM rules were, of
course, constructed so that the well established rules of classical
mechanics would be recovered as a limiting case.  The continuous
equations of motion in CM correspond to the limit in which Planck's
``fundamental grain size in phase space'' goes to zero.  In this
limit, a finite-sized physical system has an {\em infinite} number of
distinguishable states and passes through an infinite subset of these
states in a finite amount of time.  Even though an actual finite-sized
physical system has only a finite number of distinct states and a
finite rate of state change, if these finite numbers are large enough
(as they would normally be in a macroscopic system) the motion is well
approximated by the continuous CM equations.

\subsection{Classical finite-state models}

The finite-state and finite-rate character of physical systems make
classical finite state models of fundamental interest in physics.
This is because classical models are easier to understand and
simulate than quantum models.

In statistical mechanics, classical finite-state lattice systems have
a long history as models of phase change phenomena in magnetic
materials and mixtures of fluids
\cite{lee,ising-experiment,decorated,monte-carlo}.  Such {\em
classical lattice gases} can be considered special cases of general
quantum lattice systems and described and analyzed using the same
quantum formalism~\cite{ruelle-book}.  By looking at these special
cases, we gain insight into macroscopic classical concepts such as
entropy and phase change by seeing them arise from underlying
classical finite-state systems.

We might hope to similarly gain insight by studying classical special
cases of quantum {\em dynamics}.\footnote{Although the goal is very
different, some work that seeks to describe {\em all} physical
dynamics as classical also focuses on situations where QM acts
classically~\cite{fredkin-five,fredkin-acm,thooft-equiv,thooft-math,wolfram}.}
Any such classical special case must (like any other finite physical
system) necessarily have only a finite number of distinct states.

The most common classical finite-state models of physical dynamics are
numerical approximations of continuum models.  These are not good
candidates to be special cases of microscopic physical dynamics,
though, because in this case it is the theoretical continuum model
that has realistic physical characteristics, not its truncated and
rounded-off finite-state realization~\cite{toffoli-pde}.  The best
candidates for exact special cases are lattice dynamics
generalizations of classical lattice gases.  These are lattice models
with exact finite-state conservation laws, that at large scales can
exhibit not only realistic thermodynamic behavior but also realistic
classical mechanical behavior such as wave propagation and fluid
dynamics~\cite{rothman-book,crystalline}.\footnote{Discrete classical
  models of general relativity~\cite{discrete-gr}, developed for use
  in quantizing gravity, might be regarded as a step towards
  describing gravity as a classical special case of finite-state
  physical dynamics.}  These models have been used to simulate
physical dynamics~\cite{marg-prl} but have not previously been studied
as classical special cases of quantum dynamics.

\subsection{Finite-state physical dynamics}

Any invertible classical logic operation can be implemented as a QM
unitary dynamics, and so can any composition of such operations.
Since there are additional operations needed to implement an arbitrary
QM unitary dynamics~\cite{elementary-gates}, classical finite-state
dynamics is logically a proper subset of quantum dynamics.

Before the advent of invertible models of
computation~\cite{bennett-rev,fredkin-bbm} and QM models of
computation~\cite{benioff,marg-qc,bennett-info} it wasn't obvious
that, from a logical standpoint, a classical dynamics could be
regarded as a special case of quantum dynamics.  Historically, the
viewpoint had been that classical dynamics is a purely macroscopic
concept, relevant only in a limit in which quantum effects become
negligible.  In retrospect the confusion stems from not considering
the possibility of invertible classical finite-state dynamics---an
understandable omission since QM was developed before computer ideas
were prevalent, let alone the idea of invertible computation!

We are glossing over a significant issue, though, when we picture
classical finite-state dynamics as a composition of unitary
operations.  In a quantum
computer~\cite{elementary-gates,bennett-info}, the structures that
switch from one unitary operation to the next are assumed to be
macroscopic and classical.  Thus it seems that our picture is
incomplete since there are unspecified things going on to implement
the explicitly time-dependent sequence of unitary operations.  It is
helpful to note, though, that if we take the limit of an infinitely
large array of finite-state elements going through a simple cycle of
unitary operations, we can recover the familiar time-independent
equations of continuum
QM~\cite{dirac-weyl,kauff,meyer,bmb-simqm,yepez}.  Thus explicit time
dependence at the microscopic scale does not preclude compositions of
unitary operations from being equivalent to more traditional
time-independent models of physical dynamics.

Even if there is nothing essential missing from explicitly
time-dependent QM models, it's not clear that they embody what we're
looking for here: a classical-quantum correspondence that is so direct
that we might hope to interpret the quantities and concepts of quantum
dynamics in a simpler classical context.  If we have to traverse a
continuum limit in order to make contact with the familiar
time-independent quantities and concepts of continuous dynamics, we
lose this directness.

\subsection{Continuous dynamics}\label{sec.commutator}

Much of the simplicity of QM derives from the use of continuous
operators.  For example, if two QM operators $A$ and $B$ obey a
commutation relationship of the form $[A,B] = c$ for $c$ a non-zero
constant, then the operators $A$ and $B$ must necessarily be
infinite-dimensional.\footnote{If $A$ is a finite dimensional
Hermitian operator with eigenstates $\ket{a_i}$ and eigenvalues $a_i$,
and similarly for $B$, then $c = \bra{a_i}AB-BA\ket{a_i} = (a_i -
a_i)\bra{a_i}B\ket{a_i}=0$.}  For this reason the fundamental
operators describing position and momentum in QM are continuous,
just as they are in CM.

When a (finite-state) physical system is described as a quantum system
in continuous space, specifying the state of the system for a finite
number of points in space completely determines the future
evolution.\footnote{A finite-sized physical system with finite energy
is effectively finite-dimensional, and the superposition of a finite
number of momentum eigenstates has only a finite number of independent
values in space~\cite{bandlimited} (see also
Section~\ref{sec.bandlimited}).}  This suggests that we may be able to
construct continuous CM models that are special cases of continuous QM
dynamics if they also have this property.

A famous example of a CM system of this sort is Fredkin's billard ball
model of computation (BBM)\cite{fredkin-bbm}.  This is an embedding of
finite-state reversible computation into continuous CM.  At integer
times, all billiard balls are found at integer coordinates in the
plane, all moving at the same speed in one of four directions.
Collisions between billiard balls perform conditional logic---where a
ball ends up after a collision depends on whether or not another ball
was present to collide with it.\footnote{This is an idealized model
and is used here only to map finite state computation into continuous
CM and QM language.  We are not concerned with issues of stability
related to using only a set of measure zero of the possible initial
states of a continuous system.}

\subsection{Finitary classical mechanics}\label{sec.finitary}

A finite-sized BBM system is an example of what we might call a {\em
finitary} CM dynamics: one for which exact future states can be
computed indefinitely using a finite number of logical operations.

Finitary CM systems are special cases of CM that are equivalent to a
finite state dynamics when observed only at integer times.  In such
systems the motion remains perfectly continuous between integer times,
but dynamical parameters (masses, spring constants, etc.) are chosen
carefully and all dynamics is assumed to be perfect so that, as long
as the initial state is chosen from a finite set of allowed states,
one of the same allowed set is seen at each integer time (we only
consider synchronous systems here, cf.~\cite{marg-qc}).\footnote{If,
in the BBM, we want the allowed set of states to include all possible
states with balls at integer positions moving at allowed velocities,
then we need to modify the CM dynamics so that collisions (e.g., head
on) that would take us off the integer grid do not occur---balls see
zero potential in these cases and just pass straight through each
other.}

Since constraints are placed only on the values of dynamical
parameters and the initial values of state variables, finitary CM
systems map all of the continuous Lagrangian machinery and
conservations of CM onto finite-state dynamics.\footnote{Least action
principles involve continuous variations, and so there is not
necessarily an equivalent principle that just involves the states at
integer coordinates.  The existence of such integer principles in
single-speed systems (such as the ones analysed in this paper) would
not, however, contradict a recent no-go theorem~\cite{bmb-nogo}.  For
models that have a macroscopic CM limit (such as the interacting
examples we discuss), integer coordinate variations become
infinitesimal variations in the limit.}  At integer times the system
is equivalent to a finite-state lattice dynamics such as a lattice
gas automaton or partitioning
automaton~\cite{rothman-book,crystalline}.  From a continuum
perspective, though, state variables arrive at integer coordinates in
space at integer times---they don't remain on a fixed lattice at
intermediate times.  This is a subtle but important distinction.  The
same distinction in QM models allows infinite-dimensional position and
momentum operators to be applied to particles that are always found at
integer coordinates in continuous space at integer times.

\subsection{Outline of the paper}

The remainder of this paper is divided into four sections.

{\em Section~\ref{sec.shift}: Quantum concepts in a classical context}
discusses simple examples of recasting a classical finite-state
dynamics as a continuous finitary CM dynamics and thence as a
continuous quantum dynamics.  The quantum description is closely
related to classical sampling theory.  Fermions and Bosons appear in
the quantum description of classically identical particles.

{\em Section~\ref{sec.kinetic-energy}: Ideal energy} uses the
correspondence between QM and finitary CM to define an ideal energy
for finitary CM dynamics.  This is the minimum total energy that is
compatible with the rate of local state change.  A partitioning
description of dynamics defines local state change.

{\em Section~\ref{sec.waves}: An elastic string model} and {\em
  Section~\ref{sec.cb}: A colliding ball model} define and analyze
finite-state partitioning models that can be embedded in a continuous
relativistic dynamics.  We discuss the relationship between ideal
energy and relativistic kinetic energy in these models.

\myfign{register}{ \hfill \mbox{%
\includegraphics[width=3in]{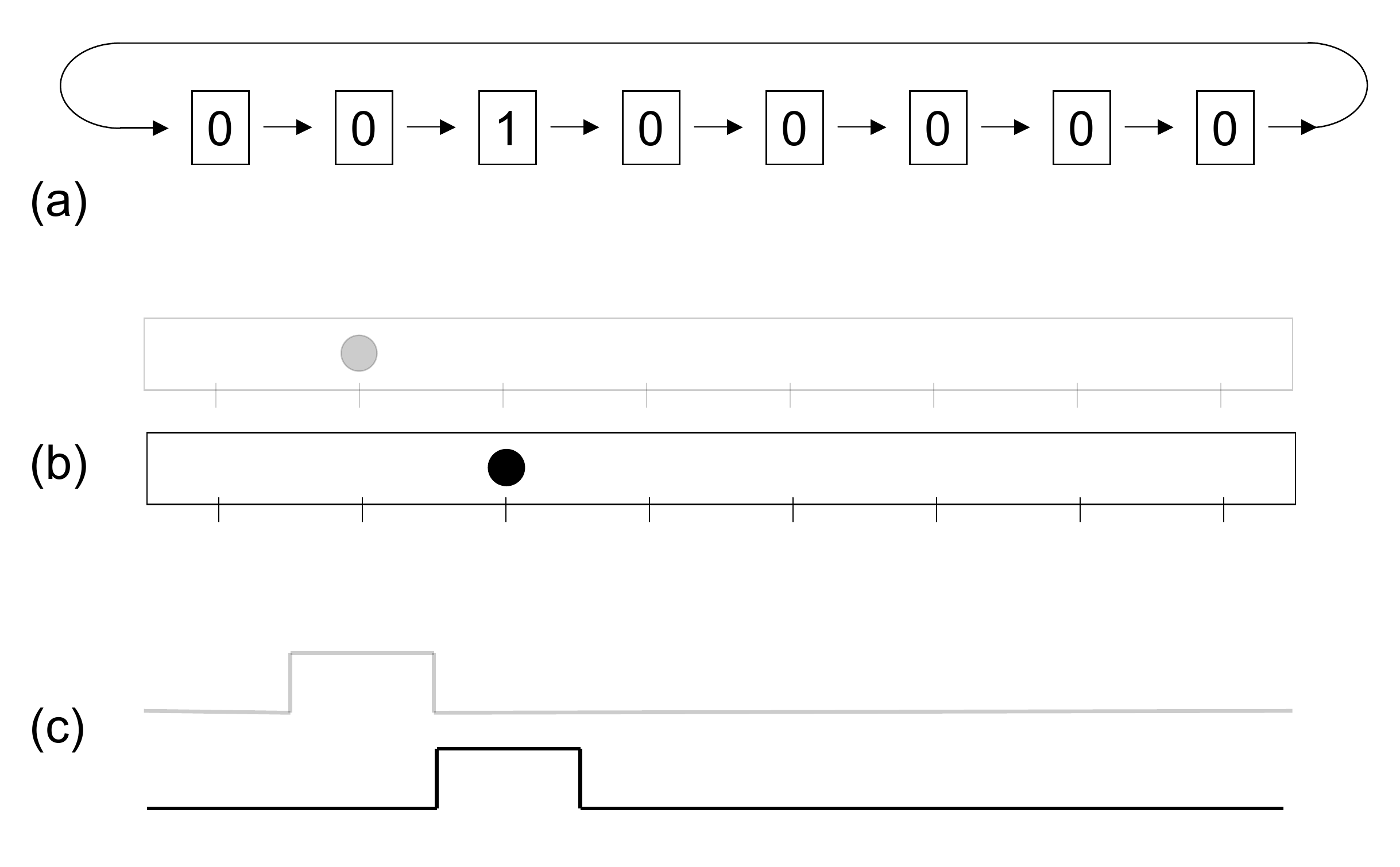}}\hfill}%
{Embedding a
finite-state classical dynamics. (a) Shift register with a single
1. (b) Classical particle in a box.  (c) Quantum particle in a box.}

\section{Quantum concepts \\ in a classical context}\label{sec.shift}

To illustrate some issues involved in describing CM dynamics using QM,
and how QM concepts look in this classical context, we discuss in this
section the simplest examples: non-interacting particle models.  These
models are embeddings of an $N$ bit shift register.  We start with the
case where all bits of the shift register are 0's except for a single
1 (Figure~\ref{fig.register}a).  At each discrete time step the
register is circularly shifted one position to the right: the 1 hops
to the next position to the right, and if it is in the rightmost
position it hops to the first position.

\subsection{Classical particle in a box}\label{sec.shift-cm}

The discrete-shift dynamics can be embedded into a continuous shift
(Figure~\ref{fig.register}b).  We replace the register with a
one-dimensional space, and the 1 by a free particle moving to the
right at a constant speed $s$.  The space is periodic, so that when
the particle exits at one end, it reappears at the other end.  If the
length of the space is $L$, then we take our unit of distance to be
$L/N$ and our unit of time to be $L/Ns$.  The particle is started at an
integer coordinate, moving to the right at speed $s$.  It will
subsequently always be found at one of the $N$ integer coordinates at
any integer time.  Interpreting the presence or absence of a particle
at a position as a 1 or a 0, the continuous system reproduces the
discrete-shift dynamics at integer times.

Conversely, the time evolution of the continuous system can be exactly
predicted at integer times by evolving the discrete-shift system.
This implies that the continuous system, with a finite set of allowed
integer-time states, effectively has no more distinct states than the
finite-state system.  From the point of view of the discrete-shift
dynamics intermediate states are {\em fictitious} states that have
been added in order to construct a continuous embedding.

In this example, as in the other finitary CM examples we will
consider, the finite subset of the continuous state set that can be
visited by the dynamics at integer times is known when the system is
defined and doesn't change with time.  Periodic discreteness makes the
details of the fictitious states between integer times irrelevant
for predicting the long-term behavior of the dynamics.

\subsection{Particle in a quantum box}\label{sec.shift-qm}

Given finite energy a QM particle in a box can be completely described
using only a finite set of distinct states.  One could regard the CM
embedding of the shift register example above as a semiclassical
analog of a QM particle in a box, illustrating how a system with a
dynamics that obeys a continuous translational symmetry can be made to
act like a finite state system by restricting the set of allowed
initial states in a manner that is conserved at integer times by the
dynamics.

We can make this more than an analogy by recasting the continuous CM
``particle in a box'' dynamics in QM language.

\subsubsection{Unidirectional Hamiltonian}\label{sec.hamiltonian}

Consider a free massless relativistic particle in a one dimensional
periodic box.  If we ignore for a moment
that the particle can travel both left and right we can take the
Hamiltonian to be $H = cp$, where $p=-i\hbar\partial/\partial x$ is
the usual momentum operator and $c$ is the speed of light.  The
Schr\"odinger equation becomes
\begin{equation}\label{eq.schrodinger}
i\hbar {\partial \psi \over \partial t} = H\psi=-ci\hbar {\partial
\psi \over \partial x}
\end{equation}
This has the solution $\psi(x,t)=f(x-ct)$, where $f(x)$ is the initial
state of $\psi$ at $t=0$ and can be specified
arbitrarily.\footnote{The close connection between classical and
quantum for this Hamiltonian is discussed in the context of hidden
variables theories in~\cite{thooft-free-rel}.  The embedding of
classical EM into a QM description in~\cite{birula-photon} is also
closely related.}  This dynamics just shifts whatever state we start
with uniformly to the right at speed $c$.

Now we assume that this system has finite energy.  In QM, this
assumption imposes constraints on the allowed initial states.  The
conventional way to analyze this for a QM particle in a box is to
describe the particle's position in terms of momentum eigenstates:
only waves that fit an integer number of wavelengths across the width
of the box are allowed by the boundary conditions.  Since momentum
eigenstates with higher spatial frequencies correspond to higher
energies, a finite maximum energy eigenstate only allows us to use a
finite number of frequency components.

\subsubsection{Energy based counting of states}\label{sec.energy-counting}

The QM average energy corresponds to the classical energy, and so the
natural constraint for a closed classical dynamical system is that the
{\em average energy} is bounded (rather than the maximum energy
eigenvalue).  We will take the average energy constraint as the
starting point for the analysis in this section.

Average energy is related to rate of state change~\cite{max-speed}:
for any closed QM system, the average energy $E$ (taking the ground
state energy as zero) gives an achievable bound on the
rate\footnote{It is easy to see that orthogonality (or approximate
orthogonality) must always occur at regular intervals, since the inner
product of any two states $\ket{\psi_t}=U_t\ket{\psi_0}$ and
$\ket{\psi_{t+s}} =U_t\ket{\psi_s}$ separated in time by an interval
$s$ depends only on $s$.} $\nu_\perp$ at which the system can pass
through a long sequence\footnote{The bound actually depends slightly
on the length $n$ of the total-system dynamical cycle that an
orthogonal change is part of---we omit a factor of $n/(n-1)$ on the
right side of the inequality.} of distinct (i.e., mutually orthogonal)
states:
\begin{equation}\label{eq.max-rate}
\nu_\perp \le 2E/h
\end{equation}
Now suppose the particle is initially very well localized.  Then the
initial wavefunction $f(x)$ is sharply peaked at one point and zero
everywhere else.  Under the ``shift'' dynamics, after a short time the
sharp peak will have moved enough so that the old and new position
states will be distinct (Figure~\ref{fig.register}c).  The more
sharply the particle is localized, the sooner the states will be
distinct and so the higher the rate of orthogonal change and hence the
higher the average energy.  If instead the ``peak'' is spread evenly
over the whole box, the shift will never produce a different state and
the average energy can be very low.

Given a finite average energy $E$ there is thus a limit to how sharply
the particle can be localized and this determines how many distinct
position states are available to the system.  If the width of the box
is $L$ and there are $N$ distinct position states, then the system
transitions to a new orthogonal state in the time it takes the pattern
to shift a distance of $L/N$.  Thus the rate at which the system
passes through orthogonal states is $\nu_\perp=cN/L$ and so, from
Equation~\ref{eq.max-rate}, $N$ can't be more than $2LE/ch$.  The
maximum number of distinct states is bounded by the energy $E$.  Since
the average (classical) momentum is $E/c$, this can also be written as
$N \le 2pL/h$.

\subsubsection{Bidirectional Hamiltonian}\label{sec.bidirect}

The unidirectional Hamiltonian we've used here seems unphysical, since
a real particle should be able to move in either direction.  The
direction should be part of the state, not part of the Hamiltonian.
This can easily be fixed.  Since the Hamiltonian $H=cp$ moves the
particle uniformly to the right at speed $c$ without changing the
shape of the wavefunction, clearly the Hamiltonian $H=-cp$ would carry
the particle uniformly to the left at speed $c$.  If we make a two
component state vector and use the Hamiltonian
\begin{equation}
H = \left( \begin{array}{cc}
   cp &  0 \\
   0  & -cp 
    \end{array} \right)
\end{equation}
then the particle will travel to the right if we represent it in the
first component, and to the left if we represent it in the second
component.\footnote{The energy associated with this Hamiltonian is
  always positive as long as we use only positive momentum eigenvalues
  to describe motion to the right at speed $c$, and only negative
  momentum eigenvalues for motion to the left.}  If we define
\begin{equation}
\sigma = \left( \begin{array}{cc}
          1 &  0 \\
          0 & -1
    \end{array} \right)
\end{equation}
then the Hamiltonian can be written as $H=c\,\sigma p$.  Thus $\sigma$
can be interpreted as the operator that reads the sign of the particle
direction from the state information.  This is a one-dimensional
version of the Weyl equation, sometimes also referred to as the
massless Dirac equation.\footnote{Lattice models that reproduce the
Weyl or Dirac equation in the continuum limit are discussed in
~\cite{dirac-weyl,kauff,meyer,bmb-simqm,yepez}.  Note that here we do
the opposite: we use the Weyl equation to exactly simulate discrete
lattice dynamics at integer times.}

\myfig{sinc}{%
\begin{array}{c@{\hspace{.6in}}c}
\includegraphics[height=1.5in]{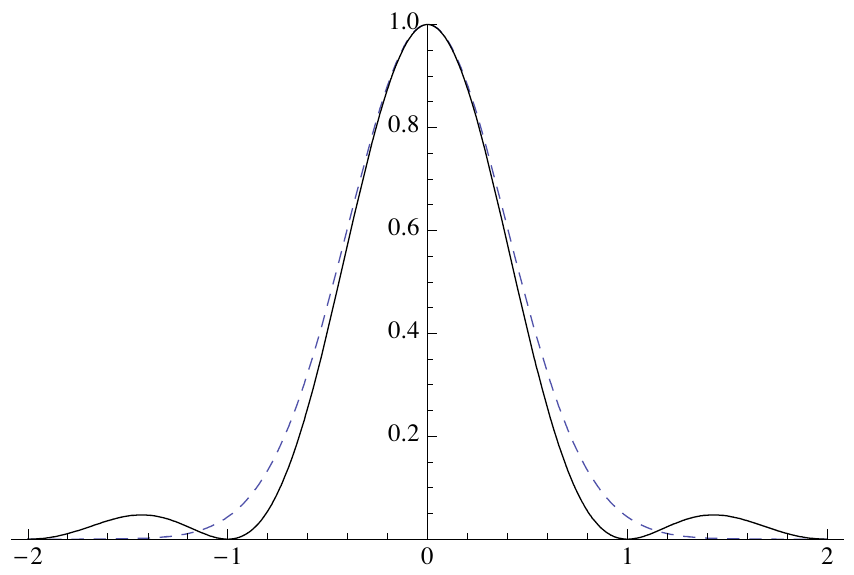} &
\includegraphics[height=1.5in]{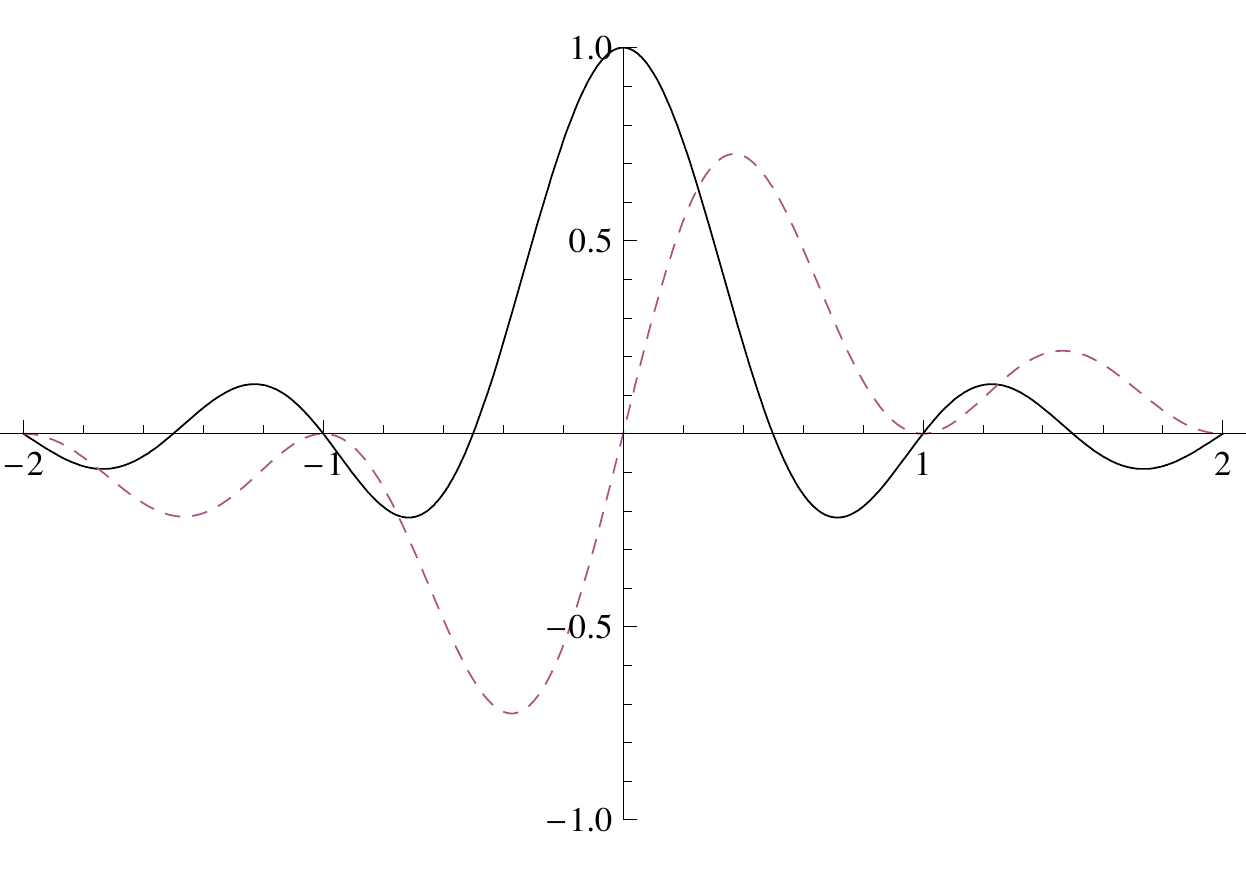} \\ 
\mbox{\bf (a)} & \mbox{\bf (b)} \\
\end{array}}
{Cardinal Sine Probability Distribution.  (a) Normalized $\sincs$
  distribution function (solid) versus normalized Gaussian of the same
  height (dotted).  (b) Real (solid) and imaginary (dotted) components
  of sinc-based amplitude distribution.}

\subsection{Discrete position basis}\label{sec.ubasis}

In Figure~\ref{fig.register}c, we depicted the wavepacket representing
the particle as a square wave.  In this section we discuss the actual
shape of the wavepacket needed to achieve the energy bound of
Section~\ref{sec.energy-counting} on the number of distinct states.

If we were just specifying a discrete distribution at integer times
and positions, the probabilities needed to reproduce the shift
dynamics are obvious: the particle is at some integer position with
probability 1 and at other integer positions with probability 0.  We
want to extend this distribution to the continuum.

For continuous positions, the width of the wavepacket represents
uncertainty in where exactly the particle would be found if we looked.
Since the shape of the wavepacket doesn't change as it shifts, we are
free to interpret all of the uncertainty as uncertainty in the initial
position of the particle.

There is, however, a better way to interpret this.  The shape of the
wavepacket is chosen to allow us to represent exactly $N$ distinct
states, each with the wavepacket centered at an integer coordinate.
The uncertainty in particle position blurs distinctions, combining
continuum states that would otherwise be distinct.  Thus another way
to interpret the position uncertainty would be to say that we invented
fictitious intermediate positions in order to map a finite-state
dynamics onto the continuum, and then we treated the exact value of
the position coordinate {\em as if it was uncertain} in order to
eliminate the fictitious position information by representing it
probabilistically as missing information.

Now, if we represented the particle by a square pulse, as in
Figure~\ref{fig.register}c, we would be assigning equal probabilities
to all fictitious intermediate positions in one region and zero
elsewhere, with a sharp transition in between.  The only constraint on
the continuous wavefunction, however, is that it should exactly
represent the finite-state dynamics when sampled at integer times and
positions.  We should therefore choose the wavefunction that includes
as little information as possible about fictitious intermediate
positions (i.e., in keeping with Jaynes Maximum Entropy
Principle~\cite{info-theory}, we should choose the wavefunction that
is maximally noncommital with regard to missing information).  This
implies a bound on the Fourier spectrum of the wavefunction: spatial
frequencies that are needed only to specify details about fictitious
intermediate positions should be omitted.

In Figure~\ref{fig.sinc}a we illustrate a probability distribution
function based on the cardinal sine function,
\begin{equation}
\sinc x = {\sin x \over x}
\end{equation}
Cardinal sine gets its name as the basis of E.\ T.\ Whittaker's
cardinal interpolation function, and it plays a central role in
bandlimited sampling theory \cite{sinc,interpol}.  It is the Fourier
transform of a finite range of frequency components, all taken with
equal weight.  The probability distribution function shown with the
solid line is $\sincs{\pi u}$.  This is a normalized distribution that
is close to a Gaussian\footnote{The use of low-pass filtering to
  extend a distribution with sharply localized features into a smooth
  continuous one is intimately related to
  diffusion~\cite{image-diffusion}.} (dotted line) near the mean, has
the value 1 at the point $u=0$ and has the value 0 at every other
integer position.


The corresponding amplitude distribution for a particle centered at
integer position $m$ is $\sinc{\pi (u-m)}$ times a phase factor.  Of
course this function also has magnitude one at the center of the
distribution, and magnitude zero at all other integer positions.  We
can use a weighted sum of such functions to construct any assignment
of amplitudes to integer positions.  Figure~\ref{fig.sinc}b shows the
real and imaginary parts of the function
\begin{equation}\label{eq.ubasis}
\phi(\pi u)=e^{i\pi u} \: \sinc{\pi u}
\end{equation}
The set of functions $\phi(\pi(u-m))$ for $m$ an integer are all
mutually orthogonal: these functions form a complete basis.  We will
refer to these states as the {\em discrete position basis}.

The discrete position basis states (and any states constructed out of
them) have a bounded spatial frequency spectrum.  These basis states
answer the question we started this section with: for the shift
dynamics, they have the waveshape that achieves the maximum possible
rate of orthogonal change for a given average energy
(Equation~\ref{eq.max-rate}).  The average energy of a particle in a
discrete position basis state is $h/2\tau$, where $\tau$ is the time
it takes the particle to travel from one integer position to the next.
We derive these fastest-changing states directly from the Hamiltonian
$H=c\, p$ in the next section.

\subsection{Bandlimited dynamics}\label{sec.bandlimited}

A general construction for fastest changing states is given in
\cite{max-speed}.  These states are essentially Fourier transforms of
a finite set of energy eigenstates, which are the slowest changing
states.

In the macroscopic limit, the discrete position basis states of
Equation~\ref{eq.ubasis} are fastest changing states (for given
average energy) of a 1D particle evolving under the shift dynamics.
The case we are interested in, however, is finite and periodic.  It is
instructive to construct the exact states that achieve the energy
bound of Equation~\ref{eq.max-rate} in this case.  If we consider a
situation where the particle travels only to the right
(Equation~\ref{eq.schrodinger}), then the energy eigenstates satisfy
the equation
\begin{equation}\label{eq.energy-eigenstates}
-ci\hbar {\partial \psi \over \partial x} = E\psi
\end{equation}
We assume that our energy scale is chosen so that the smallest value
of $E$ is zero.  This equation is satisfied by an exponential of $x$.
If we use positive $x$ coordinates that range from 0 to $L$ in the
box, then the generic solution that is periodic at $x=L$ is given by
$\psi_n \propto e^{2\pi i n x/L}$, where $n$ is a non-negative
integer.  This function satisfies Equation~\ref{eq.energy-eigenstates}
provided
\begin{equation}\label{eq.en}
E = n h {c \over L}
\end{equation}

In our CM embedding of an $N$-bit shift register dynamics, the only
meaningful positions at integer times are $x_u = u\,L/N$ for $u$ an
integer: the wavefunction at the (fictitious) non-integer positions is
a continuous extension, completely determined by the values at the
integer positions.  Thus we require that the probabilities of finding
the particle at integer positions at integer times add up to one, and
so
\begin{equation}\label{eq.normalized}
\psi_n (x_u) = {1 \over \sqrt{N}} e^{2\pi i n u/N}
\end{equation}
for which $\sum_u \psi_n^* (x_u) \psi_n (x_u)=1$.  Using
Equations~\ref{eq.en} and \ref{eq.normalized} we can construct a
normalized fastest changing state centered at position zero: in this
case it is simply the sum of the $N$ different $\psi_n$'s taken with
equal amplitudes~\cite{max-speed}.  Therefore the fastest changing
state centered at position $x_m$ is given by
\begin{eqnarray}
\Phi_m (x_u) & = & {1 \over \sqrt{N}} \sum_{n=0}^{N-1}  
                  \psi_n (x_u - x_m)   
\label{eq.fastest1} \\ 
              & = & {1 \over N} \sum_{n=0}^{N-1} e^{2\pi i n (u-m) /N}
\label{eq.fastest2} \\ 
              & = & 
                  {\sin{\pi(u-m)} \over N \sin{\pi ({u-m \over N})}}
                  e^{i\pi (u-m)(1- {1\over N})
\label{eq.fastest3}}
\end{eqnarray}
where the final expression comes from summing the geometric series and
some rearrangement.  The fastest changing states are Fourier
transforms of the energy eigenstates.  For large $N$, $\Phi_m(x_u)$
turns into $\phi(\pi(u-m))$ of Equation~\ref{eq.ubasis}.
$\Phi_m(x_u)$ also has the property that it is one for $u\equiv m
\imod{N}$ and zero for $u$ any other integer.  For integer values of
$u$, the $\Phi_m(x_u)$ functions provide a complete discrete-position
basis for the finite system.

\subsection{Both continuous and discrete}\label{sec.both}

The discrete position basis state $\Phi_m(x_u)$ corresponds to a
probability distribution that is normalized both from a continuous and
a discrete point of view.  Furthermore, the discrete normalization
doesn't apply just to a sampling of the state at integer locations: it
applies to any discrete sampling of the state at a unit-spaced set of
positions.

The continuous normalization is apparent from
Equation~\ref{eq.fastest2}, since
\begin{eqnarray}
\displaystyle\int_0^L dx \; \Phi_0^* (x) \, \Phi_0 (x)
\hspace{-1in}
\nonumber \\
& = &
{1 \over N^2} \sum_{n=0}^{N-1}\sum_{n'=0}^{N-1} \int_0^N du \;
e^{2\pi i u (n'-n)/N} \nonumber \\
& = & 1
\end{eqnarray}

To see the generality of the discrete normalization, we similarly sum
over positions $x_{u+\delta}$ where $u$ is an integer and $0 < \delta
<1$.  This shifted normalization also implies that if we start the
system in a discrete-position basis state, then as the wavefunction
shifts continuously with time the values sampled at fixed integer
locations always constitute a normalized discrete representation of
the state.

Both normalization properties are of course inherited by any
normalized wavefunction $\Psi(x,t)$ constructed out of the $\Phi_k$
states (i.e., any $\Psi$ whose spectrum of spatial frequencies is
bandlimited like the $\Phi$'s).  $\Psi(x,t)$ can be exactly
reconstructed from the amplitudes $\Psi(x_k,t)$ sampled at integer
locations:
\begin{equation}
\Psi(x,t) = \sum_{k=0}^{N-1} \Psi(x_k,t) \, \Phi_k (x)  \label{eq.reconstruct}
\end{equation}

\subsection{Intermediate positions}\label{sec.moving-basis}

We can construct a $\Phi_c$ state centered at any continuum position
$c$ as a superposition of the $N$ discrete-position basis states
$\Phi_m$ centered at integer locations $m$.  This is obvious since the
$\Phi_m$ states form a complete basis for describing the continuously
shifting time evolution of an initial $\Phi_m$ state.  A shift of all
$\Phi_m$ states by the same amount yields a new basis.

If we start our system with a particle at an integer location in a
basis state and then change bases in step with the particle motion, we
can follow the continuous motion of the particle, always seeing it in
a single localized basis state.  From this point of view the quantum
motion is just as spatially local and continuous as the classical
motion: in both cases we see the constraint on the classical system of
a discrete set of possible positions at integer times reflected as a
corresponding discrete set of possible intermediate positions at each
intermediate time.

\subsection{Classical Fermions and Bosons}\label{sec.fermions}

The shift register example is easily extended to multiple particles.
This simply corresponds to a value in the register
(Figure~\ref{fig.register}a) with more than a single `1' bit.  For
simplicity, we'll assume for the moment that all particles are
traveling to the right.

In elementary QM, the overall state of a collection of independent
particles is normally represented using a product of single-particle
states: we track the state of each particle separately as if it were the
only particle in the system.

For our classical shift register we've already described a single
particle system using a localized wavepacket basis
(Equation~\ref{eq.fastest3}).  Here we'll denote the single particle
basis state where particle $j$ has its wavepacket centered at location
$x$ as $\ket{x,j}$.  An overall state can be described as a product of
such states, one per particle.

Since all particles are moving to the right, the direction $\sigma$ is
+1 in all cases.  The Hamiltonian for a multi-particle system is then
$H=\sum_j cp_j$, where $p_j$ acts only on particle $j$.  For example,
for a two particle system (i.e., shift register with two 1's), the
Schr\"odinger equation would be
\begin{eqnarray}
i\hbar {\partial \ket{x,0}\ket{y,1} \over \partial t} & = &(cp_0 +
cp_1)\ket{x,0}\ket{y,1} \nonumber \\ 
& = & (cp_0 \ket{x,0})(cp_1 \ket{y,1})
\end{eqnarray}
This equation is satisfied if both $i\hbar {\partial \ket{x,0} /
\partial t} =  cp_0 \ket{x,0}$ and $i\hbar {\partial \ket{y,1} /
\partial t} =  cp_1 \ket{y,1}$ hold, and so the two particles each
follow a shift dynamics.

By using a product of single-particle wavefunctions to construct our
multiparticle wavefunction we've added two kinds of extra states that
didn't exist in the original shift register.  First of all, our states
can represent more than a single 1 at the same location, which isn't
possible in the original shift register.  Secondly, our states keep
track of which particle is which, but these particles are identical
1's in a shift register and interchanging two 1's in a binary number
doesn't give us a different state!

We can fix both of these problems by combining each set of equivalent
states into a single state.  We do this by adding together all product
states that differ only by a relabeling of identical particles and we
call that a single state.  In performing this sum on a set of $M$
equivalent states we weight half of the states by a factor of
$+1/\sqrt{M}$ and half by $-1/\sqrt{M}$.  This is done in such a
manner that the entire combination changes sign if we interchange any
two particle labels (i.e., it is {\em antisymmetrized}).  In our two
particle example, the antisymmetrized wavefunction is
\begin{equation}
\psi = {\ket{x,0}\ket{y,1} - \ket{y,0}\ket{x,1} \over \sqrt{2}}
\end{equation}
Using only antisymmetrized wavefunctions we can no longer represent
two particles (1's) at the same location: as in this example, if two
mean positions are equal (i.e., $x=y$) the wavefunction is zero.

Replacing each set of equivalent states with a single antisymmetrized
sum avoids over-representing states.  For a classical embedding into
QM, a superposition of classically distinct states corresponds to a
classical statistical ensemble: we can interpret all probabilities as
arising from ignorance about which single state the system started in.
Thus by assigning equal probabilities to a set of equivalent states,
the antisymmetrized superposition represents the equivalent states as
if we were completely ignorant about which particle was which.  Of
course the particle label information doesn't exist in the original
finite-state dynamics and so, once again, the probabilities we've
assigned represent ignorance of fictitious classical details added in
constructing our QM description.

Thus for a classical dynamics of bits, if a fixed number of identical
1's are described quantum mechanically as if we could distinguish
which is which, and then the fictitious distinctions are eliminated
using superposition, they are {\em Fermions}.\footnote{Deterministic
  Fermions have been described previously in a formal QM
  field-theoretic context (e.g.,~\cite{thooft-equiv}).}  If, instead,
there were no constraint in our original classical dynamics on how
many 1's can be represented at each location, we would use
symmetrization rather than antisymmetrization to eliminate the
fictitious distinctions.  In this case the 1's would be {\em Bosons}.

\subsection{Non-classical amplitudes}

There are two kinds of probabilities that can arise in a quantum
description of a classical finite-state dynamics.  There may be
unknown information about the state of the actual dynamics: this is
represented by ordinary probabilities and contributes to the entropy
of the system.  There may also, however, be unknown information about
fictitious classical details that were added in constructing the
quantum description of the dynamics (such as distinct labels for
identical 1's): this is represented by probabilities derived from
amplitudes and doesn't contribute to the entropy of the system.

There is more to amplitudes than this, though.  If the dynamics was
always described in a single classical basis, then the superposition
would evolve like a classical statistical ensemble and the coefficient
of a basis state in the superposition could be any function of the
basis state's ensemble probability---there is no recombining of
coefficients in the ensemble dynamics.  The fact that amplitudes are
square roots of probabilities makes it possible to also use them to
describe the system in multiple bases: the sum of the squares of
component magnitudes is the same for a vector described in any basis.
Amplitudes, when used for basis change, still don't represent unknown
information about the real system, and so again don't contribute to
its entropy.

\subsection{Finite-state unitary dynamics}

For a classical finite-state system with a conserved number of
identical 1's, we can treat the 1's as distinguishable particles in a
QM description and follow their trajectories.  We can then use
appropriate symmetrization of superpositions of products of single
particle states to merge equivalent states that aren't actually
distinguishable.

We can extend this approach to describe classical systems in which the
numbers of 1's and 0's are not constant.  The creation (and
annihilation) operators of quantum field theory allow us to act on an
appropriately symmetrized superposition of products of single-particle
states, to add (or remove) a single-particle basis state in each
product term while maintaining the symmetrization of the overall
state.  We can use an occupation-number basis to avoid dealing
directly with single-particle states, but a vestige of
antisymmetrization persists as anticommutation of Fermionic field
operators~\cite{ziman}.

We can avoid all of this complexity entirely, though, by simply not
describing identical 1's as if they were distinguishable!  We won't
have to nullify fictitious particle labels using symmetrization if we
don't put them in in the first place.  Similarly, unnecessary
complexity is added when we employ a continuum of positions and times
to describe a classical system with only a finite number of possible
states.

Both sources of complexity are eliminated if the QM dynamics is
constructed to be isomorphic to the classical finite-state dynamics.
If we start with a classical dynamics expressed as a composition of
invertible logic operations we can, in the QM description, simply
replace each classical logic operation with an equivalent finite-time
finite-state unitary transformation.  There are then no extra states
added to the description that need to be nullified.  Moreover, if the
finite-state dynamics can be interpreted as an integer-time sampling
of a finitary CM dynamics, then it is easy to identify the familiar
continuous-space and -time conserved quantities of CM, including
classical energy.  The corresponding identification of QM energy is
discussed in the next section.

\section{Ideal energy}\label{sec.kinetic-energy}

In CM there is no minimum energy required to allow a given sequence of
state changes to take place in a given period of time.  In QM, there
is.  If a classical finite-state dynamics is recast isomorphically as
a unitary QM dynamics we can place a bound on the least possible
average energy for physically realizing the classical dynamics---the
{\em ideal energy} of the dynamics (cf. \cite{marg-qc}).

For a dynamics described as a composition of spatially-localized
invertible operations, Equation~\ref{eq.max-rate} allows us to define
a minimum energy bound $E_\statechange$ that depends on counting state
changes:
\begin{equation}\label{eq.bce}
E_\statechange = {h \over 2\tau}* (\mbox{state-changes in $\tau$})
\end{equation}
Here $\tau$ is the length of the time interval during which we count
state changes.  Each invertible operation that actually changes the
state counts as one state change.

Now, in CM, energy of motion appears as localized state change.  Thus
the state-change energy we define here can be thought of as a kind of
kinetic energy: it bounds classical energy of motion, at a given time.
The total classical energy is then bounded by the maximum value that
the state-change energy acquires in the course of the time evolution.

In this section we discuss state-change energy and use it to define
the ideal minimum energy of a finitary CM dynamics.  In
Sections~\ref{sec.waves} and \ref{sec.cb} we apply this idea to two
examples of finitary CM dynamics.

\subsection{Classical partitioning dynamics}\label{sec.partitioning}

If a spatially extended finite-state system evolves in time
synchronously, the rate of global state change dramatically
underestimates the minimum energy needed for an actual physical
realization of the dynamics.  In particular, the energy of a
collection of independent subsystems is additive: localized
independent changes must be counted separately.

We can arrive at a more realistic minimum-energy estimate by making
use of the locality and time dependence inherent in a physical
description of classical finite state dynamics.  Consider, for
example, a finitary CM dynamics such as Fredkin's BBM.  In this model,
all collisions happen synchronously, at a discrete set of possible
locations at a discrete set of possible times.  If we focus on a time
interval starting shortly before a possible collision time and ending
shortly after, then during this interval the system can be partitioned
into a set of disjoint regions, each of which evolves independently:
within each region one or more particles head towards the locus of a
possible collision, collide or don't collide, and then head away from
the locus.  Which regions evolve independently must of course change
with time or the particles could never cross between regions.  Thus
the partitioning used in a description of the system as a collection
of independent subsystems must be time dependent, even though the
continuous dynamics itself follows a time independent dynamical law.

\myfign{balls}{%
\begin{array}{c}
\includegraphics[height=1.5in]{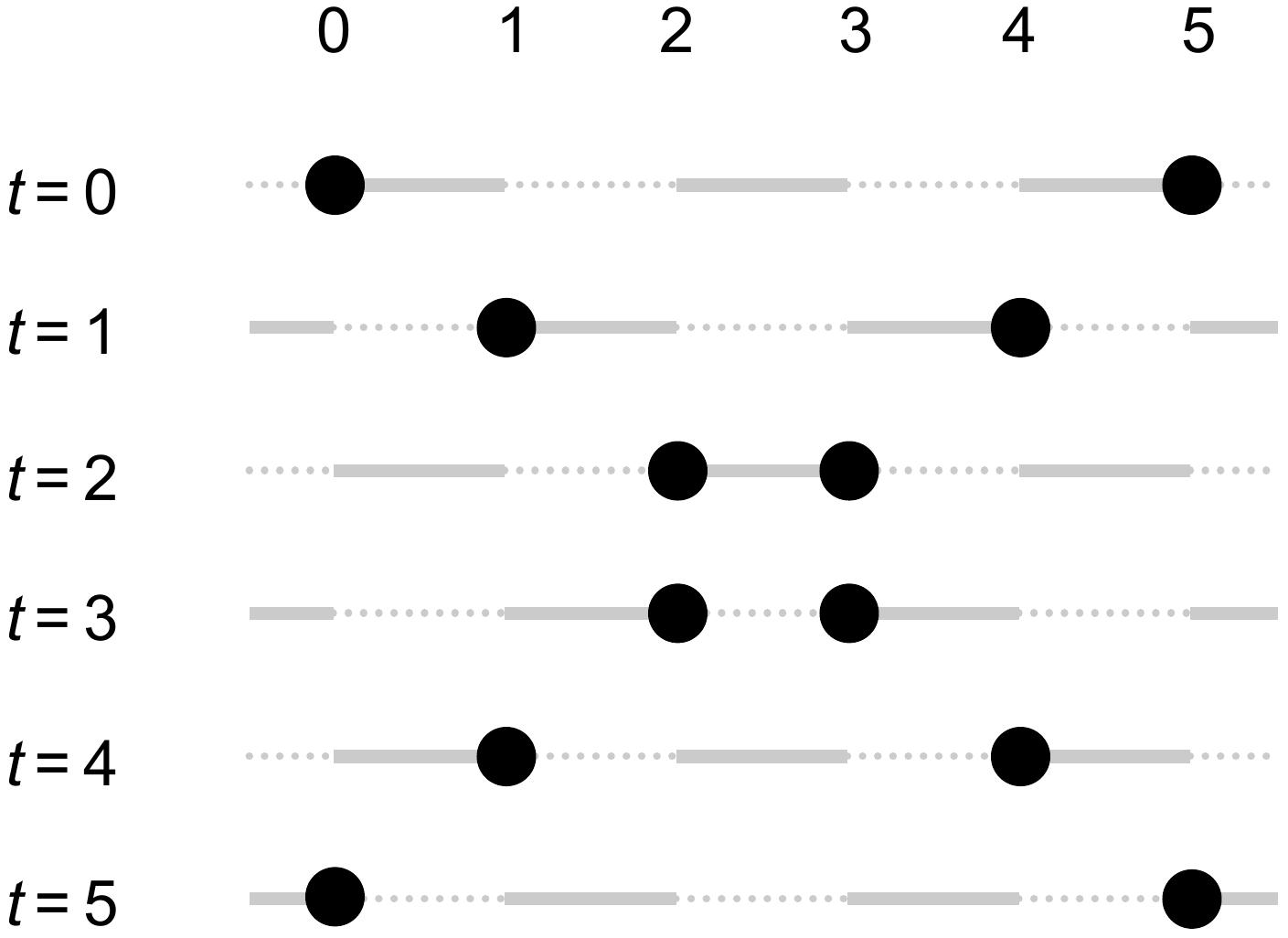} \\
\end{array}}
{Integer-time history of two particles moving continuously in a
  periodic box.}

Figure~\ref{fig.balls} illustrates the partitioning idea applied to
the ``classical particles in a periodic box'' model discussed in
Section~\ref{sec.shift}.  A time history from this finitary CM
dynamics is shown, with the positions of two continuously moving
particles (1's) sampled at integer times.  If the spatial intervals
marked with solid lines in the figure are taken to be blocks of a
partition, then during the period between one integer time and the
next any particles that start off in a block remain in that block and
only interact with other particles in the same block.  The classical
finite-state dynamics at integer times can be summarized as: swap the
values (1's or 0's) at the two integer locations in each block of the
partition.

Notice that, looking just at the integer-time history, it is ambiguous
whether continuously moving particles whose paths intersect pass
through each other or bounce back.  This kind of ambiguity is very
common in extending a finite-state dynamics to construct an equivalent
finitary CM dynamics.  If, however, we only want to quantify the
minimum possible amount of motion consistent with a given classical
finite-state dynamics, then there is no ambiguity.  For example, for
the dynamics illustrated in Figure~\ref{fig.balls}, the least amount
of motion occurs if the two particles don't move between steps 2 and
3.

Of course in a real CM dynamics, the amount of motion that occurs
inside the block where the two particles collide and bounce back could
be small if they slow down quickly, but not exactly zero.  We might
think of zero motion either as a limit or, alternatively, as the exact
amount of motion seen at the midpoint of the block update.


\subsection{Ideal kinetic energy}

The integer-time {\em swap dynamics} of Figure~\ref{fig.balls} can be
described as a time-dependent sequence of unitary operations, with
operations acting first on each block of one partition, and then on
each block of the other partition.

Equation~\ref{eq.bce} assigns $h/2\tau$ of state-change energy to each
block update involving a single moving particle.  This agrees with our
earlier QM analysis of the minimum achievable energy for a freely
moving CM particle.  Since there is no kinetic energy associated with
blocks that don't change, either in the CM or the QM description, the
state-change energy defines an ideal kinetic energy in this case: it
is at all times the minimum kinetic energy that could possibly be
achieved in a physical realization of the dynamics.  Its maximum value
is the ideal total energy.

In general, the definition of kinetic energy in a finitary CM dynamics
is somewhat ambiguous.  For example, the same dynamics might be
interpretable as either a relativistic system or a non-relativistic
system~\cite{soft-sphere}.  If the state-change energy is taken as the
ideal kinetic energy at all times, this ambiguity is resolved.  Even
if this is not done, however, the state-change energy sets the ideal
energy scale, since it bounds the minimum {\em total} energy that can
be involved in each possible block transition: for any CM
interpretation of the dynamics one of these bounds will be the most
constraining.

Below, we discuss ideal energy in two example models, beginning with
a model that is closely related to the swap dynamics.

\myfig{waves}{%
\begin{array}{c@{\hspace{.6in}}c@{\hspace{.6in}}c}
\includegraphics[width=1.7in]{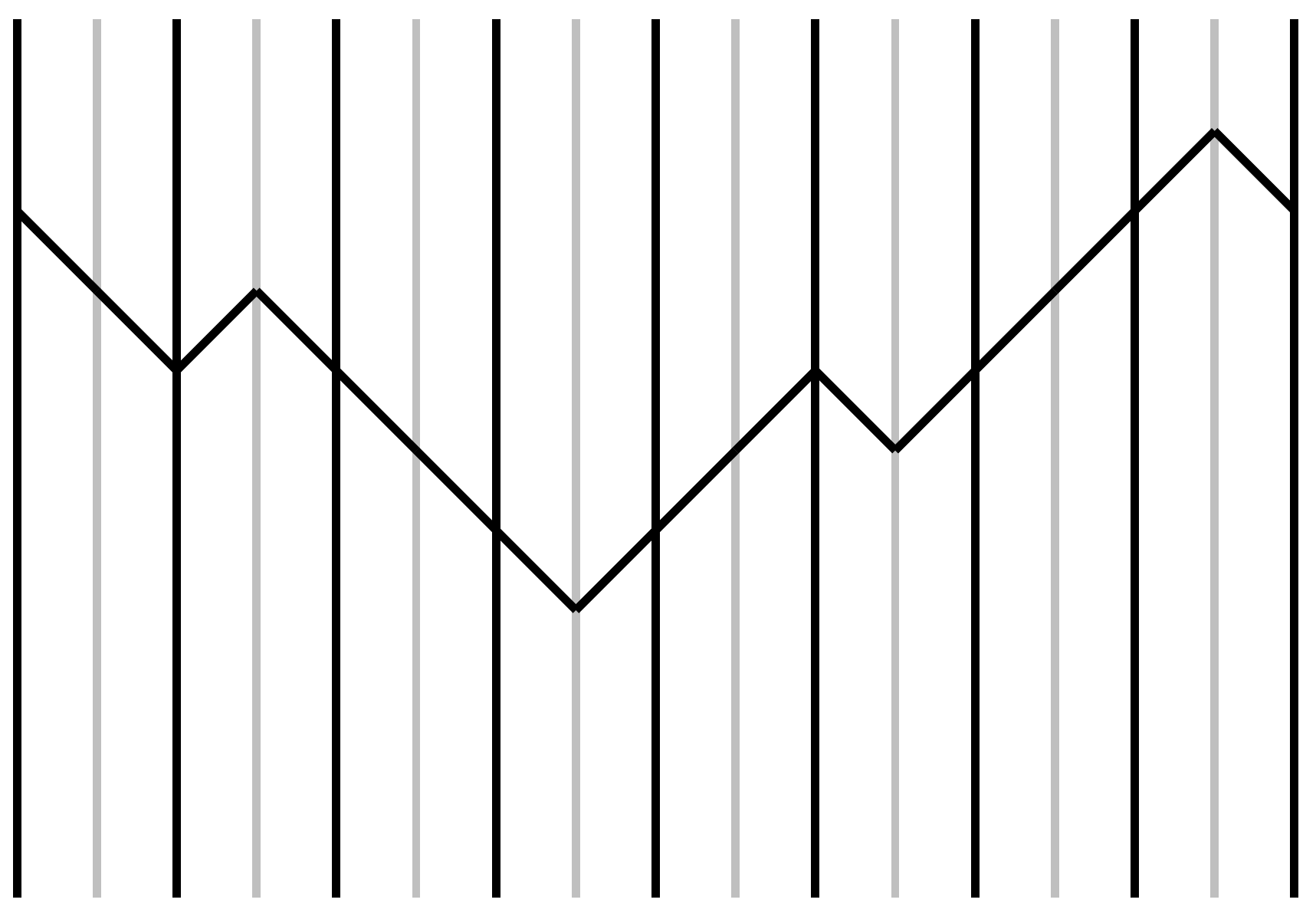} &
\includegraphics[width=1.7in]{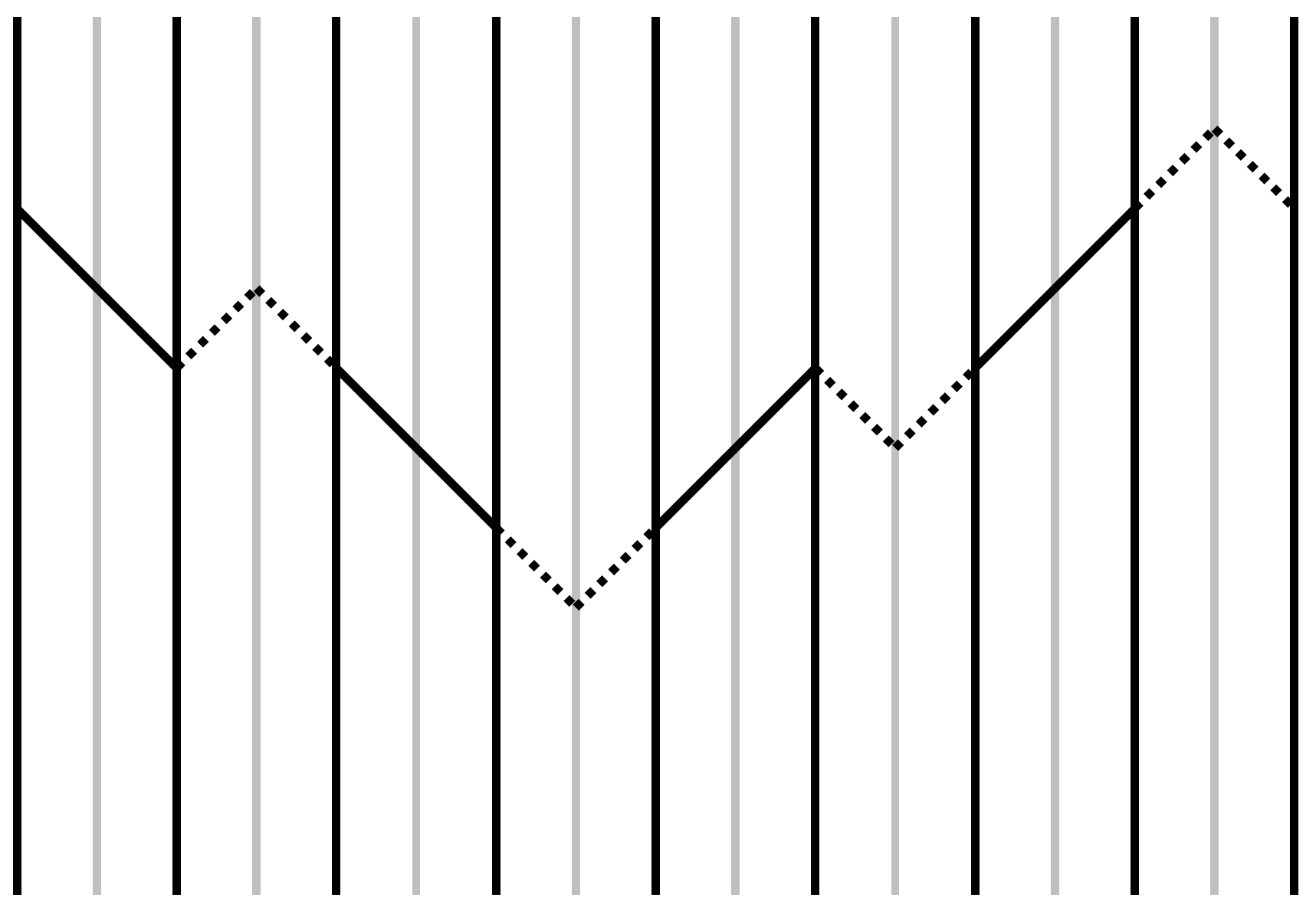} &
\includegraphics[width=1.7in]{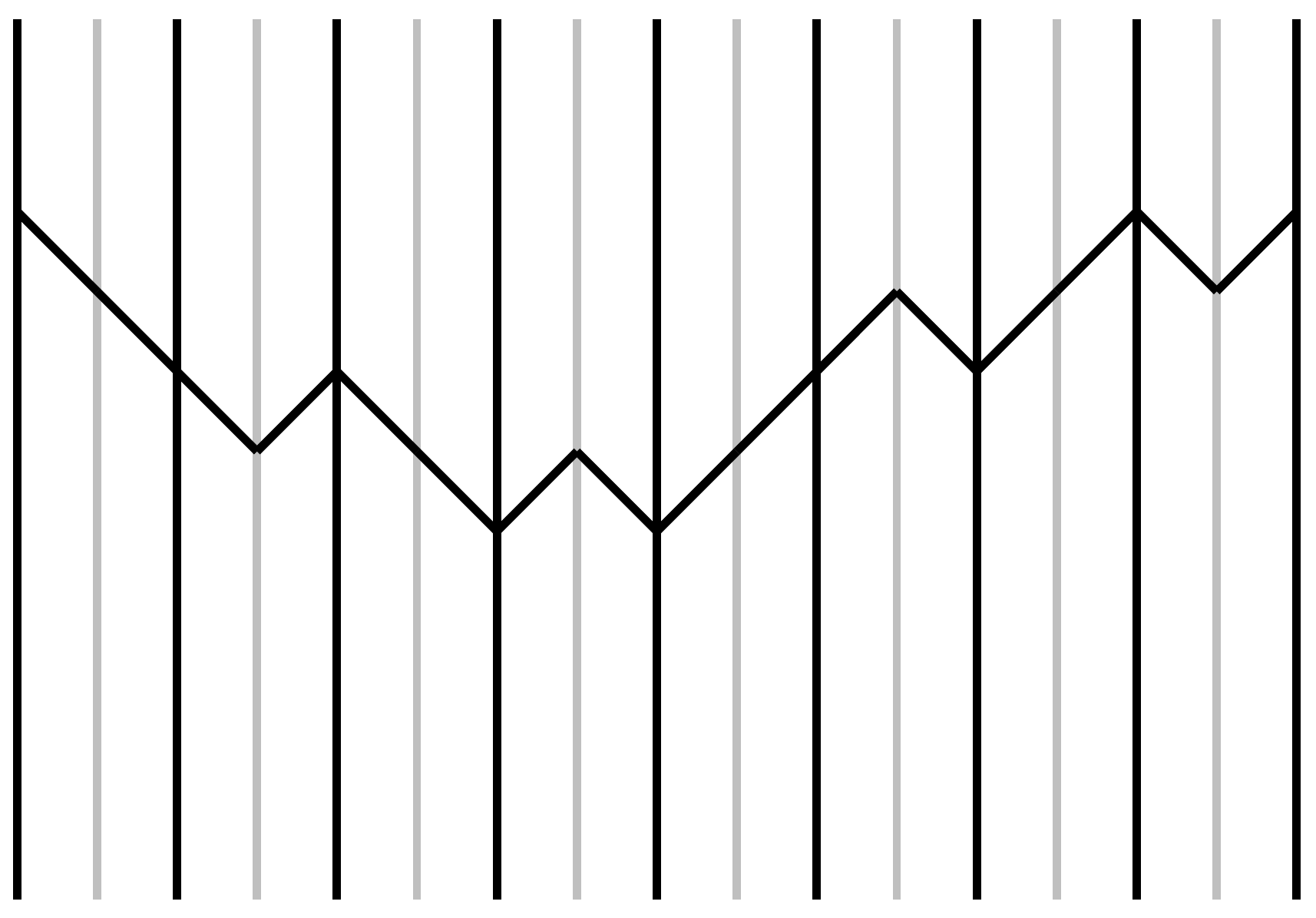} \\ 
\mbox{\bf (a)} & \mbox{\bf (b)} & \mbox{\bf (c)} \\


\end{array}}
{Discrete wave dynamics. Elastic string is held fixed where it crosses
  black bars.}

\section{An elastic string model}\label{sec.waves}

In this section we discuss a classical finite state model of wave
motion in an elastic string.  The finite-state string model
corresponds exactly to the integer-time behavior of a continuous CM
model of the type discussed in Section~\ref{sec.finitary}.  The
finite-state string model also {\em exactly} reproduces the behavior
of the time-independent one-dimensional wave equation sampled at
integer-times and locations.  There is a finite-state constraint on
the allowed initial shape and motion of each unit segment of the wave
and this constraint reappears at each integer time.  In the continuum
limit the discrete constraint on the wave disappears; the exactness of
the wave dynamics itself (at discrete times) is not dependent on this
limit.

The model presented here has been discussed
before~\cite{hh,crystalline,toffoli-action}, but the analysis of
translational motion, the relativistic interpretation and the analysis
of ideal-energy given here are all new.

\subsection{Discrete wave model}

Consider an ideal continuum string for which transverse displacements
exactly obey the wave equation.  In Figure~\ref{fig.waves}a we've
illustrated an initial configuration with this string stretched
between equally spaced vertical bars.  The set of initial
configurations we're allowing are periodic, so the two endpoints must
be at the same height.  Any configuration is allowed as long as each
segment running between vertical bars is straight and lies at an angle
of $45^\circ$.

Initially the string is attached at a fixed position wherever it
crosses a vertical bar.  We start the dynamics by releasing the
attachment constraint at all of the gray bars.  The attachment to the
black bars remains fixed.  In Figure~\ref{fig.waves}b the segments
that are about to move are shown with dotted lines: the straight
segments have no tendency to move.  Under continuum wave dynamics, the
dotted segments all invert after some time interval $\tau$.  This will
be our unit of time for the discrete dynamics.  The new configuration
at the end of this interval is shown in Figure~\ref{fig.waves}c.  At
this instant in time all points of the string are again at rest and we
are again in an allowed initial configuration.  At this instant we
interchange the roles of the black and gray bars and allow the
segments between adjacent gray bars to move for a time inteval $\tau$.
The dynamics proceeds like this, interchanging the roles of the black
and gray bars after each interval of length $\tau$.  Since attachments
are always changed at instants when all energy is potential and the
string is not moving, the explicit time dependence of the system
doesn't affect energy conservation.

\myfig{waves-rule}{%
\begin{array}{c@{\hspace{.6in}}c@{\hspace{.6in}}c}
\includegraphics[height=1.5in]{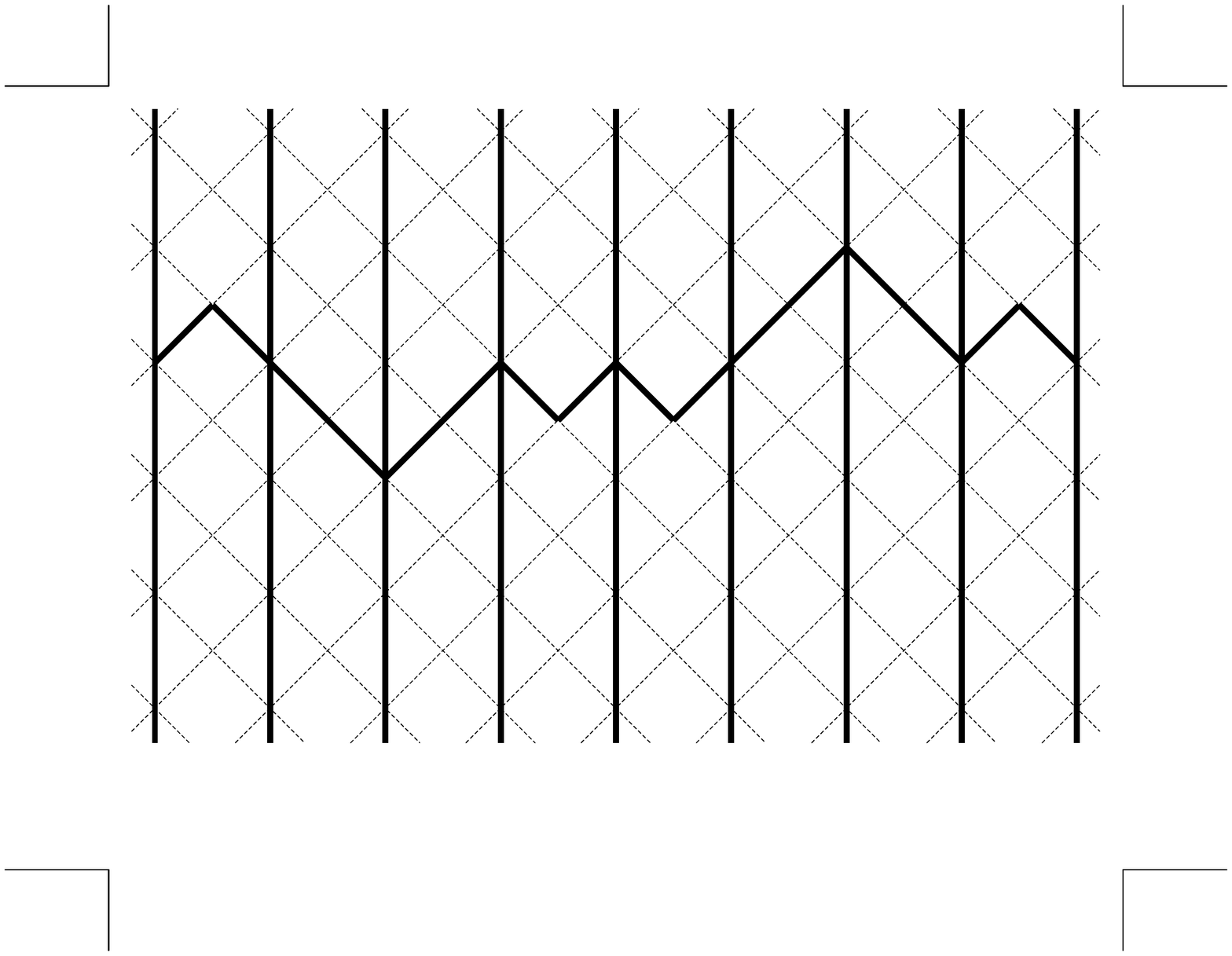} &
\includegraphics[height=1.5in]{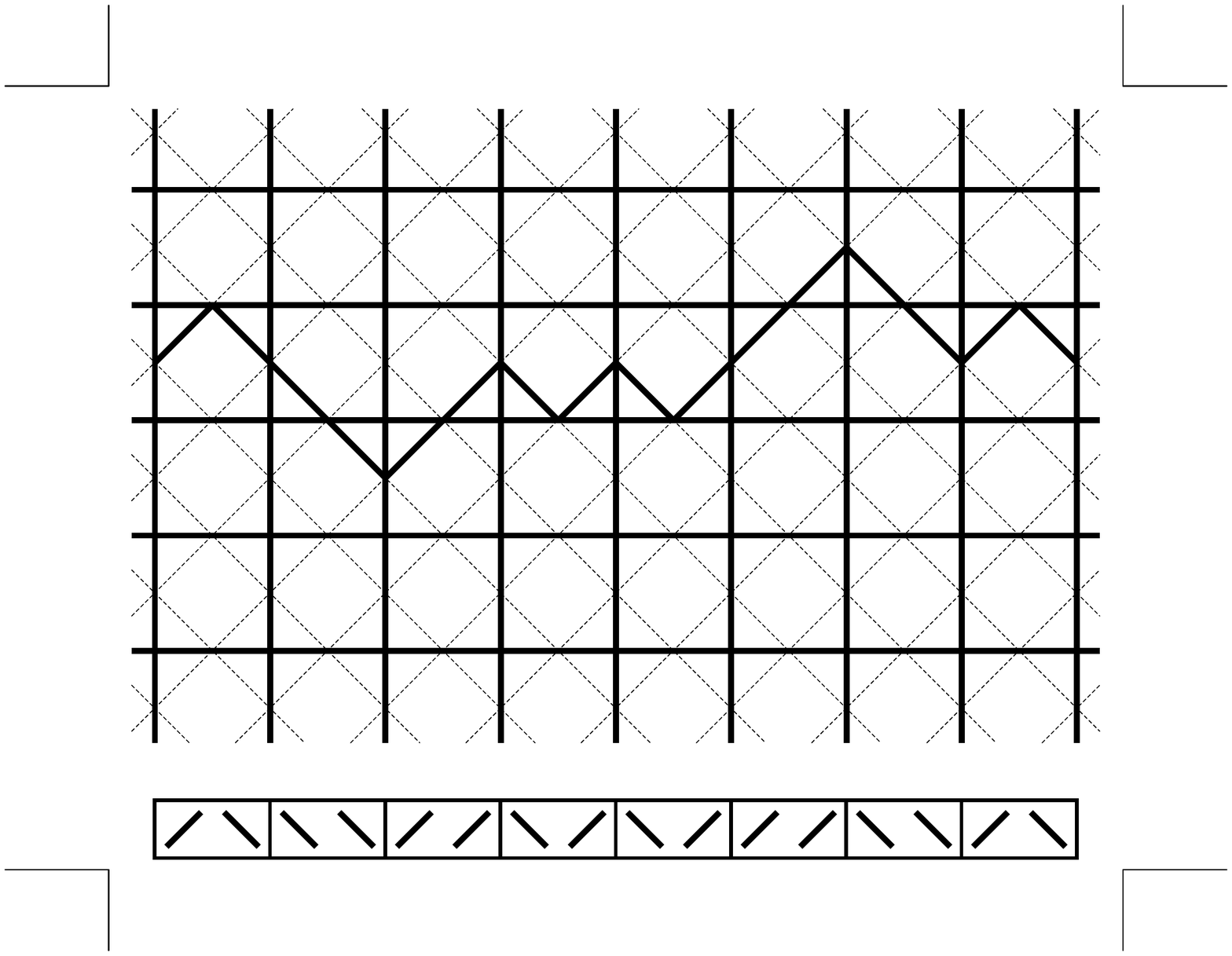} &
\includegraphics[height=1.5in]{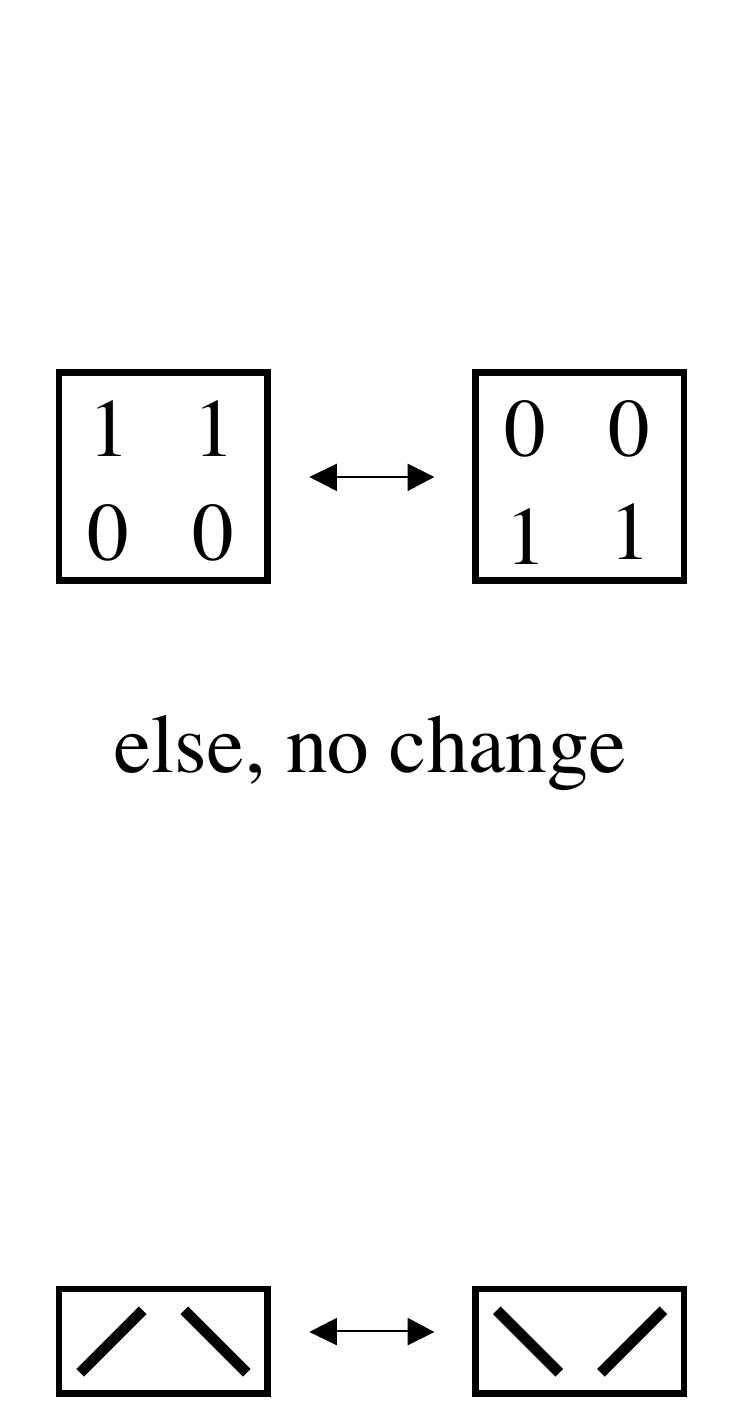} \\ 
\mbox{\bf (a)} & \mbox{\bf (b)} & \mbox{\bf (c)} \\ 

\end{array}}
{Discrete wave dynamics.  (a) Wave configuration with possible wave
  paths indicated (dotted lines).  (b) One of two partitions used for
  a discrete update rule (horizontal and vertical lines).  Array of
  wave gradients shown at bottom.  (c) Top, dynamical rule for wave.
  Presence of wave-path segments is indicated by 1's.  Bottom,
  equivalent dynamical rule for gradients.}

We express this dynamics as a purely digital rule in
Figure~\ref{fig.waves-rule}.  In Figure~\ref{fig.waves-rule}a we show
the wave just after the evolution from the previous figure, with the
black bars shown marking the attachments for the next step.  To
simplify the figure we have suppressed the gray bars---they are always
situated midway between the black bars and so don't need to be shown.
We have also added a grid of $45^\circ$ dotted lines that shows all of
the segments that the string could possibly follow.  In
Figure~\ref{fig.waves-rule}b we add in horizontal black bars, in order
to partition the 2D grid into a set of 2$\times$2 blocks that can be
updated independently.  Note that in all cases the segments that are
allowed to change during this update step, as well as the cells that
they will occupy after the update, are enclosed in a single block.
Segments that aren't going to change stretch across multiple blocks.
The 1D box below Figure~\ref{fig.waves-rule}b contains just the slope
information from the string.  This array of gradients is clearly
sufficient to recreate the wave pattern if the height at one position
is known.

Figure~\ref{fig.waves-rule}c shows the dynamical rule for a block.
Since the dotted lines indicate the direction in which segments must
run if they appear in any cell, the state information for each segment
is only whether it is there or not: this is indicated with a 0 or a 1.
The only segments that change are peaks \verb|/\| or valleys
\verb|\/|, and these are represented by two 1's at the top of a block
or at the bottom of a block respectively.  The rule is that peaks and
valleys turn into each other, and nothing else changes.  This rule
applies to the blocking shown, and then to the complementary blocking
when the attachments change, alternating back and forth.  This is a
partitioning dynamics.

\myfig{scale-wave}{%
\begin{array}{c@{\hspace{.8in}}c@{\hspace{.8in}}c}
\includegraphics[height=1.4in]{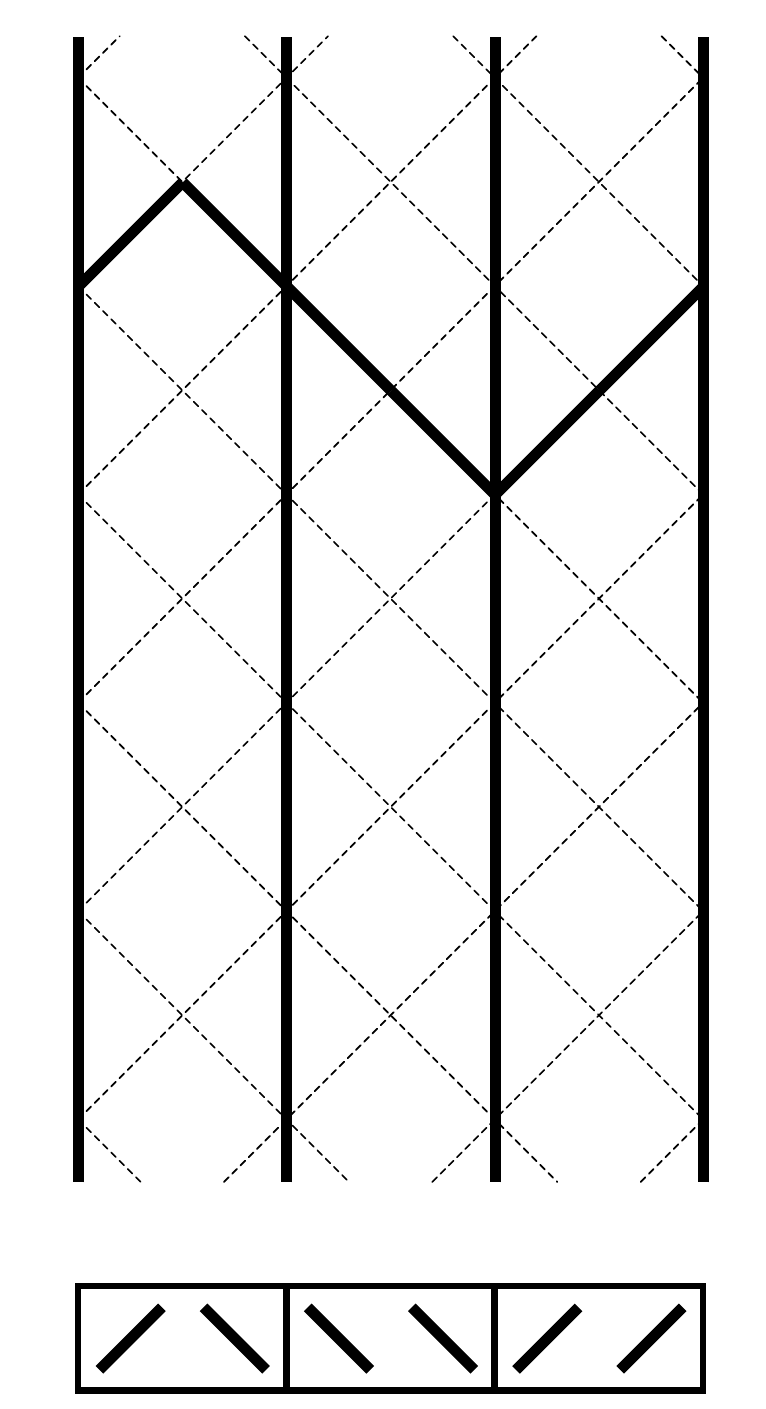} &
\includegraphics[height=1.4in]{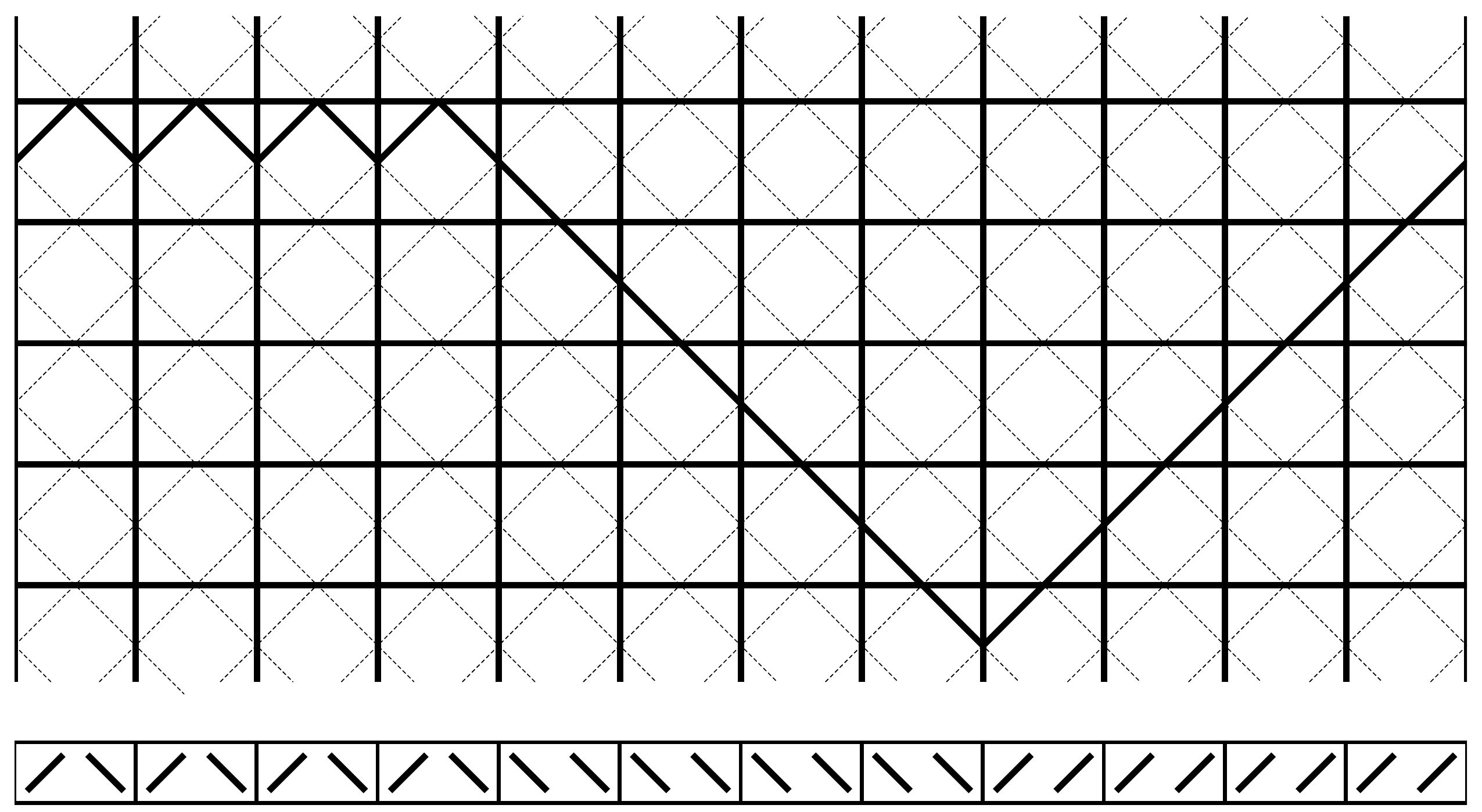} &
\includegraphics[width=1.2in]{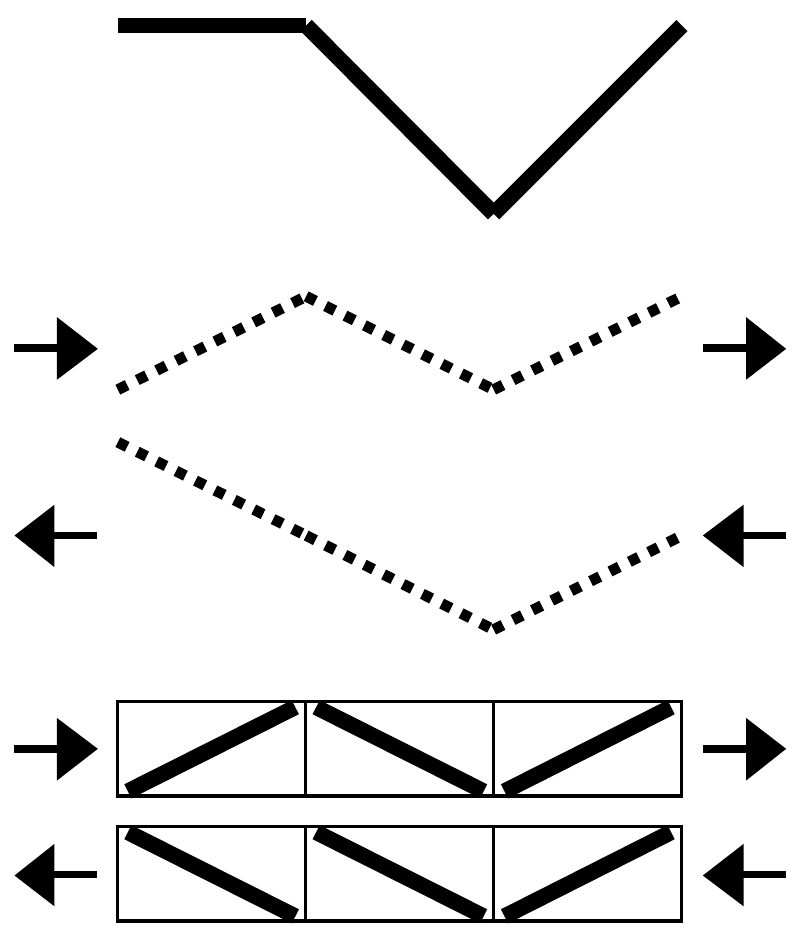} \\ 
\mbox{\bf (a)} & \mbox{\bf (b)} & \mbox{\bf (c)} \\ 
\end{array}}
{Rescaling a discrete wave.  (a) Original wave and array listing its
  gradients.  (b) Each pair of gradients is repeated four times.  The
  corresponding wave is drawn by placing gradients end-to-end.  (c)
  Equivalent superposition of continuous waves.}

\subsection{Exact wave behavior}

At the bottom of Figure~\ref{fig.waves-rule}c we've presented a
dynamics for the {\em gradient} of the wave.  Since the full 2D
dynamics just turns peaks into valleys and vice versa, leaving
straight segments unchanged, we can perform that dynamics equally well
on the array of gradients!  As the 2D dynamics interchanges which
blocking to use, the dynamics on the gradients also alternates which
pairs of gradients to update together.  In all cases, the dynamics on
the gradients duplicates what happens on the string: if the two
dynamics are both performed in parallel, the gradient listed below a
column will always match the slope of the string in that column.

The dynamics on the gradients has a very interesting property.
Turning a peak into a valley and vice versa is exactly the same as
swapping the left and right elements of a block.  Leaving a \verb|//|
or \verb|\\| unchanged is also exactly the same as swapping the left
and right elements of a block.  In all cases, the dynamics on the
gradients is equivalent to a swap.

This means that the left element of a block will get swapped into the
right position, and at the next update it will be the left element of
a new block and will get swapped into the right position, and so on.
Thus all of the gradients that start off in the left side of a block
will travel uniformly to the right, and all of the gradients that
start in the right side of a block will travel uniformly to the left.

This shows that the system obeys a discrete version of the wave
equation.  Half of the gradients constitute a right-going wave, and
half constitute a left-going wave.  At any step of the dynamics, the
2D wave in the original dynamics is just the sum of the two waves: it
is reproduced by laying gradients end to end!

As the number of cells in our lattice gets arbitrarily large, the
straight segments in our waves become smaller and smaller compared to
the total width of the lattice, and the discrete wave equation turns
smoothly into exactly the continuum wave equation.  As we will see,
though, even the discrete model exactly obeys the continuum wave
equation.

\subsection{Rescaling limit}

The discrete wave system also has another continuum limit that allows
us to embed our discrete wave dynamics with time-dependent blocking
into a continuous dynamics with a time-independent dynamical law.

Take an array of gradients that define a wave
(Figure~\ref{fig.scale-wave}a) and make a longer list by replicating
every pair of segments $N$ times, producing a list that is $N$ times
as long (Figure~\ref{fig.scale-wave}b).  Since the dynamics for the
gradients is simply that all of the even numbered gradients shift one
way and all of the odd numbered gradients shift the other, the state
after $N$ steps corresponds exactly to the state we would have gotten
by running one step on the original configuration, pairing the
gradients into the alternate set of blocks and then replicating each
block $N$ times.

Although the $N$-step dynamics is exactly equivalent to the original,
the waves we reconstruct from the array of repeated blocks look much
different than simply a scaled up version of the original wave: a peak
\verb|/\| doesn't turn into a peak $N$ times as large, it turns into
$N$ little peaks \verb|/\/\/\.../\|.  This is a wiggly flat line, and
as $N$ grows the wiggles get proportionately smaller and smaller
compared to the straight parts \verb|//| and \verb|\\|, which simply
get $N$ times longer.  In the limit, the peaks turn into flat places
with a net velocity up or down, still obeying the continuum wave
equation.

\subsection{Continuous at all scales}

The rescaled continuum configurations correspond isomorphically with
the discrete time evolution, since a flat portion of the wave moving
up or down can simply be interpreted as an array of peaks or valleys.
In fact, if we simply draw each block of the active partition that
we've been showing as \verb|/\| or \verb|\/| as a horizontal segment
moving up or down instead, then the discrete dynamics is directly
mapped onto the continuous wave equation without invoking any limiting
process!

Since this evolution obeys the continuous wave equation we can draw it
as the superposition of continuous rightgoing and leftgoing waves.
This is shown in Figure~\ref{fig.scale-wave}c.  The top solid wave is
a redrawing of Figure~\ref{fig.scale-wave}a with a horizontal segment
used instead of the \verb|/\| (or equivalently, the continuum limit of
Figure~\ref{fig.scale-wave}b).  It is also the sum of the dotted
rightgoing and leftgoing waves shown below it, each segment of which
has slope $\pm 1/2$.  The collection of gradients that make up the
rightgoing and leftgoing waves are shown at the bottom.  These are
just the even position (top) and odd position (bottom) gradients from
Figure~\ref{fig.scale-wave}a, each stretched to fill the width of a
block.  After the waves each move half the width of a gradient
segment, all segments will again be aligned and will add up to a
result that corresponds to the next step of the discrete wave
dynamics, with the appropriate blocking.

\subsection{Relativistic motion}\label{sec.rel-string}

Assume the string carrying a discrete wave wraps around a space of
width $N$.  We've discussed the horizontal motion of waves along such
a string, but the string itself can move vertically.  For example, a
pattern such as \verb|/\/\/\.../\| all the way around the space
reproduces itself after two steps, but shifted vertically by two
lattice units.  This is clearly the maximum rate of travel for a
string: one position vertically per time step.

We can express the net velocity of the string in terms of the
populations of rightgoing and leftgoing gradient segments.  With $N$
positions there are $N/2$ rightgoing segments and $N/2$ leftgoing
segments, and the members of each group never change.  Thus if there
are $R$ rightgoing \verb|\|'s, there must be $(N/2)-R$ rightgoing
\verb|/|'s, and similarly for the other direction.  Therefore if we
know the numbers $R$ and $L$ of rightgoing and leftgoing \verb|\|'s,
we have complete population information.

For a wave that wraps around the space and joins at the ends, there
must be equal numbers of \verb|/|'s and \verb|\|'s.  Thus half of the
gradient segments must be \verb|\|'s, 
\begin{equation}\label{eq.r+l}
R+L={N \over 2}
\end{equation}
The average upward velocity of the string depends on the difference
between $R$ and $L$,
\begin{equation}\label{eq.r-l}
{R-L \over N/2} = v
\end{equation}
If we think of the dynamics as swapping adjacent pairs of gradients
during each step (even though \verb|\\| and \verb|//| pairs don't
actually change), each gradient moves one position up or one position
down during each update.  A \verb|\| moving right always moves the
gradient it passes one up as it goes by, whereas a \verb|\| moving
left moves the gradient it passes one down.  Thus we get
Equation~\ref{eq.r-l} for the average velocity of gradients that pass
the $N/2$ \verb|\|'s.  The average velocity of gradients that pass the
\verb|/|'s is exactly the same (their populations are complementary
but the effects of passing them are also complementary).

Since the motion of each gradient contributes a velocity of $\pm 1$ to
the the net motion of the string, it is natural to interpret the net
motion relativistically.  We can focus on just the \verb|\| particles,
since the \verb|/| particles have the same behavior.  The total energy
is proportional to the number of identical particles all moving at the
same speed, so we let $E=R+L=N/2$ (Equation~\ref{eq.r+l}).  Since
$P=Ev$ (taking $c=1$), this gives us a vertical momentum of $P=R-L$
(Equation~\ref{eq.r-l}).  Then the mass of the string must be $M^2 =
E^2 - P^2 = 4RL$ and $E=\gamma M$ where $\gamma = (1-v^2)^{-1/2}$.

We can exactly analyze the vertical motion of the string
statistically, since the probability $p = R/(R+L)$ is the {\em exact}
frequency with which leftgoing gradients will encounter a rightgoing
\verb|\| in the course of a cycle of length $N/2$.

A rightgoing gradient is the left element of a block of two adjacent
string segments that are about to be updated by the dynamical rule.
If the block contains \verb|\/| then the dynamics will move both
elements upward one position.  Thus $p$ is the probability that, for
any block holding part of the string, the left element of the string
is ready to take an upward step, and $1-p$ the probability it is ready
to take a downward step.

Similarly, if a block contains \verb|\| as its right element (i.e., a
leftgoing \verb|\|), then the block is ready to take a downward step
if the left element is \verb|/|.  The chance that the block contains a
leftgoing \verb|\| is $L/(R+L)$ which is just $1-p$.  Thus in either
position of a block, $p$ is the probability a gradient is ready to
take an upward step and $1-p$ the probability it is ready to take a
downward step, and so the average frequency (per step) of both
elements of a block moving upward is $p^2$ and both downward is
$(1-p)^2$.  The average upward velocity of the string is thus $p^2 -
(1-p)^2$ which just gives us back
Equation~\ref{eq.r-l}.\footnote{Since $p$ is also the probability of
moving a gradient upward as it passes and $1-p$ the probability of
moving it downward, the average vertical velocity $v$ is also given by
$v=p-(1-p)$.}

\subsection{Relativistic kinetic energy}\label{sec.string-ke}

Our original continuous description of the elastic string dynamics
involved minimal-motion between integer times: all cases remain
motionless except for \verb|\/| and \verb|/\|, which turn into each
other.  Thus the least possible energy for this transition determines
the ideal kinetic energy for the string dynamics.

Since the average frequency (per step) of both elements of a block
moving upward is $p^2$ and both downward is $(1-p)^2$, the average
fraction of the blocks that change per step is $p^2 + (1-p)^2$.  This
fraction times $N/2$, the total number of blocks in the string, gives
(using Equation~\ref{eq.bce}) the least possible average kinetic
energy (in energy units of $h/2\tau$).  Since the motion exactly
repeats after a cycle of $N/2$ steps, the ideal kinetic energy per
cycle is invariant.

The minimum value of $p^2 + (1-p)^2$ is $1/2$ and occurs when $p=
1/2$, which is also when $v=0$.  Thus at low speeds, half of the block
changes don't contribute to the net vertical motion of the string.  If
we think of the string as a particle moving in one dimension (i.e.,
vertically) with a horizontal internal dimension of width $N$, then
only the portion of the ideal kinetic energy that contributes to
vertical motion is particle kinetic energy.  Thus at low speeds, the
ideal particle kinetic energy is the excess over $1/2$ of the blocks
that change,
\begin{eqnarray}
T_\nonrel   & = & {N\over 2} \left(p^2 + (1-p)^2 - {1\over 2}\right)
                  \nonumber \\
            & = & E \left( \left( {1+v \over 2} \right)^2  +
                           \left( {1-v \over 2} \right)^2
                           - {1\over 2}\right) 
                  \nonumber \\
            & = & {1 \over 2} E v^2  \label{eq.nrel-ke}
\end{eqnarray}
Since $E=\gamma m$, for small $v$ we have $E=m$ and we recover the
expected non-relativistic kinetic energy.

As $p$ approaches zero or one, the fraction of the blocks that change
approaches one and the vertical speed of the string also approaches
one.  Thus as $v \to 1$, all of the ideal kinetic energy contributes
to the string motion: $T \to E$.

\subsection{Time dilation}

If we think of the block changes that don't contribute to overall
string motion as internal dynamics of the string, then as the string
approaches the speed 1 the internal dynamics stops: all of the changes
contribute to overall string motion and none to internal dynamics.
The internal dynamics exhibits relativistic time dilation.  There is a
slight subtlety, though, that arises from relativistically
interpreting {\em all} gradient segments as moving.  We
correspondingly interpret a fraction $1/\gamma$ of {\em all} segment
motion as contributing to the internal motion, and so the internal
dynamics slows down by this factor as the string speeds up.  The
remaining fraction $1-1/\gamma$ of the segment motion contributes to
the overall string motion.  Of course some of the internal ``motion''
we're counting here doesn't involve block changes, but all of the
segment motion that contribute to overall motion of the string does actually
correspond to block changes.  Thus the fraction of all $N/2$ blocks of
the string that contribute to kinetic energy of string-particle motion
is in fact $1-1/\gamma$ (i.e., $T=E-m$).  The fraction of blocks that
contribute only to internal kinetic energy in the string is then the
total fraction that change minus the external kinetic fraction, and so
\begin{eqnarray}
T_\internal & = & {N\over 2}
     \left(p^2 + (1-p)^2 - \left(1- {1\over \gamma}\right)\right)
     \nonumber \\
            & = & E\left({1\over \gamma} - {1-v^2 \over 2}\right)
\end{eqnarray}
which is always non-negative and which approaches a fraction of all
blocks of $1/2 - O(v^4)$ as $v$ approaches zero.\footnote{Notice that
in this model, for a fixed string length $N$ the energy of the string
is fixed and the rest mass $E/\gamma$ approaches zero for strings that
move at speeds approaching 1.}

Related models of diffusive behavior which make contact with
relativity are discussed in
~\cite{smith,curious-properties,toffoli-action}.
None of these relativistic discussions define a model with objects
that have an internal dynamics.

\section{A colliding ball model}\label{sec.cb}

In this section we construct a new invertible partitioning dynamics
that has a large-scale limit with macroscopic objects and forces.
This finite-state dynamics is then analyzed as the integer-time
behavior of a continuous relativistic CM dynamics.  We use the
state-change energy bound of Equation~\ref{eq.bce} to define the ideal
energy of the model and we compare ideal kinetic energy with
relativistic kinetic energy.

\myfign{ssm}{%
\includegraphics[height=1.5in]{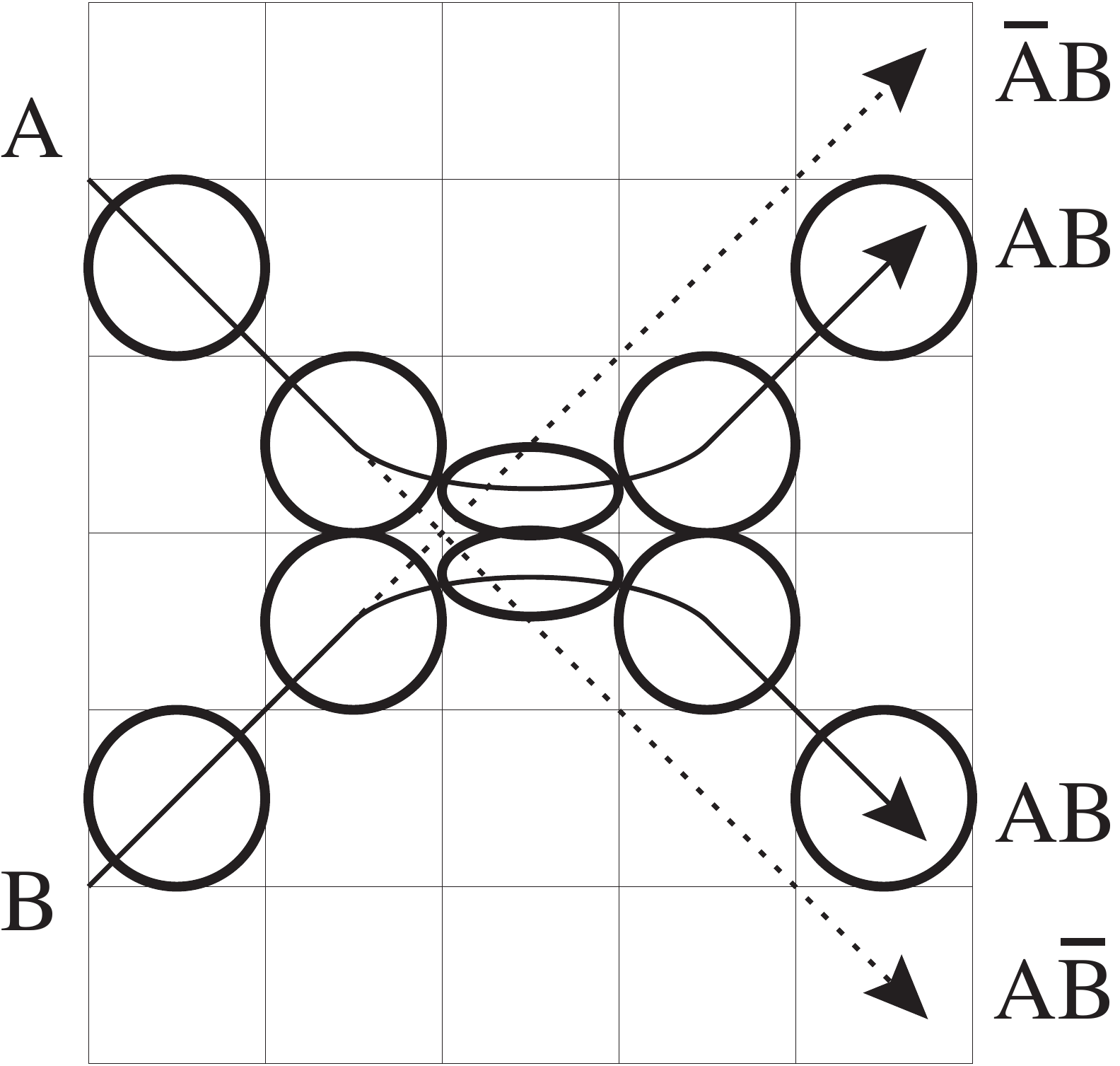}}
{A collision in the Soft Sphere Model finitary CM dynamics.}

\subsection{Soft Sphere Model}

In Figure~\ref{fig.ssm} we illustrate the soft sphere model
(SSM)~\cite{soft-sphere}.  The SSM is an invertible and energy
conserving CM model of classical computation similar to Fredkin's
billiard ball model.  Unlike the BBM, however, it can be turned into a
lattice gas dynamics~\cite{rothman-book,crystalline} with point
interactions.

As in the BBM we restrict the initial positions and velocities of
balls to discrete values, to guarantee that the SSM acts like a
digital system.  In the collision shown in
Figure~\ref{fig.ssm}, the horizontal component of velocity of
all balls entering from the left is one column per time step, and so
consecutive moments of the history of the collision occur in
consecutive columns.

The collision shown is energy and momentum conserving and the
compression and rebound takes exactly the time needed to displace the
colliding balls from their original paths onto the paths labeled $AB$.
If a ball had come in only at $A$ with no ball at $B$, it would leave
along the path labeled $A\bar{B}$.  This model is equivalent to a
lattice gas automaton, with lattice sites located at the corners of
the grid shown in the figure.

\myfign{rel-coll}{%
\begin{array}{c@{\hspace{.4in}}c}
\includegraphics[height=1.5in]{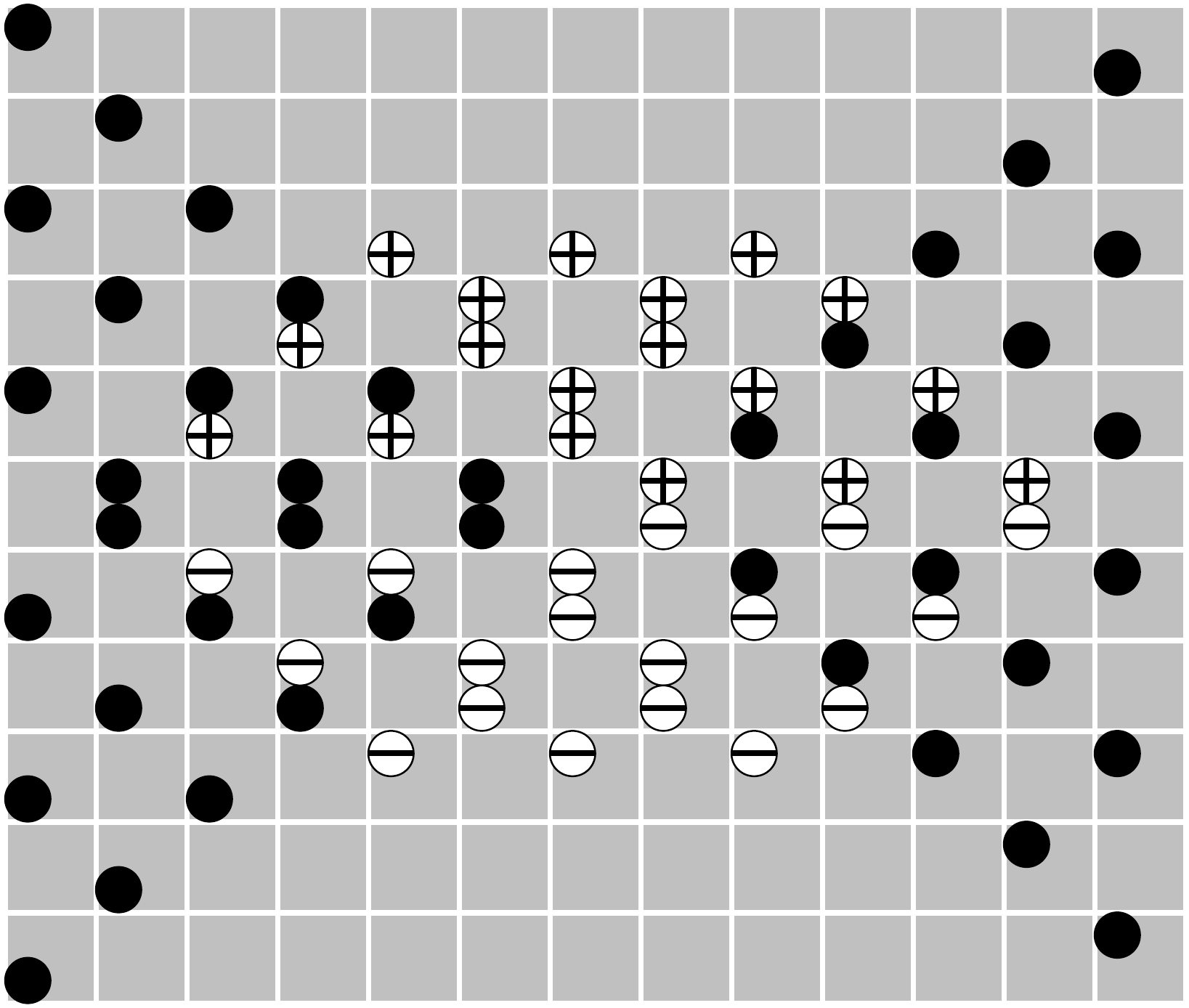} &
\includegraphics[height=1.5in]{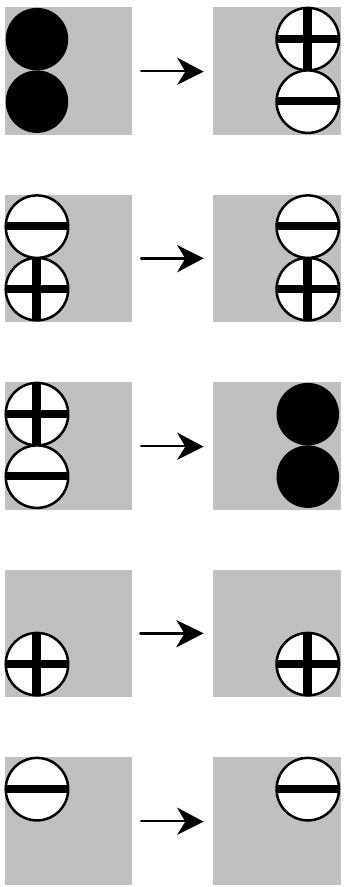} \\
\mbox{\bf (a)} & \mbox{\bf (b)} \\
\end{array}}
{Rescalable soft sphere partitioning dynamics.  (a) A time history of
  one column of particles colliding.  At the times shown all particles
  are converging towards the center of the square they are in.  (b)
  Site update rule.  After the update particles are diverging from the
  center of the square.  Only these cases interact (and rotations);
  otherwise particles go straight.}

\subsection{Rescalable SSM}\label{sec.rss}

Figure~\ref{fig.rel-coll} describes the Rescalable Soft Sphere (RSS)
model which implements a partitioning version of the SSM using
colliding blocks of 1's of any size.  This is a new invertible model
that conserves energy and momentum.  Colliding blocks undergo
collisions and then reform in appropriately shifted positions to
exactly implement the SSM dynamics on any scale.  The basic idea is
that the size of the incoming blocks controls how long the interaction
takes, because particles behave differently when they are surrounded
by other particles than when they are alone.  This allows the shift
caused by a collision to depend on the size of the colliding blocks.

Figure~\ref{fig.rel-coll}a shows a simple collision.  Two groups of
three particles approach each other (first column on the left), with
the top group moving right and down, the bottom group right and up.
As in Figure~\ref{fig.ssm}, all particles have a constant horizontal
velocity and so consecutive discrete moments in the history of the
collision occur in consecutive columns.

Each square in Figure~\ref{fig.rel-coll}a is a block of the even-step
partition and columns are shown at consecutive even times.  This means
that all particles within a square are converging towards the center
of the square at the times shown.  For example, since the top three
black particles in the first column are in upper left squares, they
are headed down and right.  To make the diagram easier to understand
we've spread out the particles in the initial state so that particles
only ever interact on even steps, at the moments shown.  The odd-step
interaction is turned off for now---particles just go straight on odd
steps.  Later we'll make both steps the same and put the particles
closer together.

The size of the incoming groups of particles controls how long the
interaction takes.  Pairs of corresponding incoming black particles
collide along the axis of symmetry between the incoming groups, each
pair colliding at a separate spot.  Each pair of colliding black
particles turns into a pair of white particles moving in the same
directions.  The white particles move straight as long as they
encounter other particles, but turn back towards the axis of the
collision as soon as they find themselves alone.  The white particles
are labeled with a chirality ($+$ or $-$) at the time they are
created, so that they will know which way to turn.  The white
particles ignore other white particles of the same chirality and don't
interact again until the $+$ and $-$ come back together at the axis of
the collision.  They then turn back into black particles and
reconstitute two groups of black particles, each of which moves off as
a unit.

\subsection{Transition rule}

The transition rule used in Figure~\ref{fig.rel-coll}a is shown in
Figure~\ref{fig.rel-coll}b.  On the left we see the particles before
the interaction, as they converge towards the center of the square.
On the right we show the particles after the interaction, as they
diverge.  In addition to the cases shown, each orthogonal rotation of
a state on the left turns into the same rotation of the corresponding
result case on the right.  This allows us to have the same collision
occur in any orthogonal orientation.  In all cases not shown, there is
no interaction.  Particles go straight through the center of the
square and continue moving in the same direction without change.

This rule can be inferred from Figure~\ref{fig.rel-coll}a.  All of the
cases shown are needed in order to reproduce the given evolution
except for the second case (white to white collision) which doesn't
occur.  This case is added to make the overall dynamics invertible.
Since cases that aren't shown don't interact, particles will often
pass through each other without affecting each other.  For example, if
two groups of black particles collide head on, they pass through each
other.  Since most of the ``no interaction'' cases don't occur in the
kind of collision shown in Figure~\ref{fig.rel-coll}a, we could
augment the rule without affecting Figure~\ref{fig.rel-coll}a by
changing some of the ``no interaction'' cases.

\myfign{rel-block}{%
\includegraphics[height=1.5in]{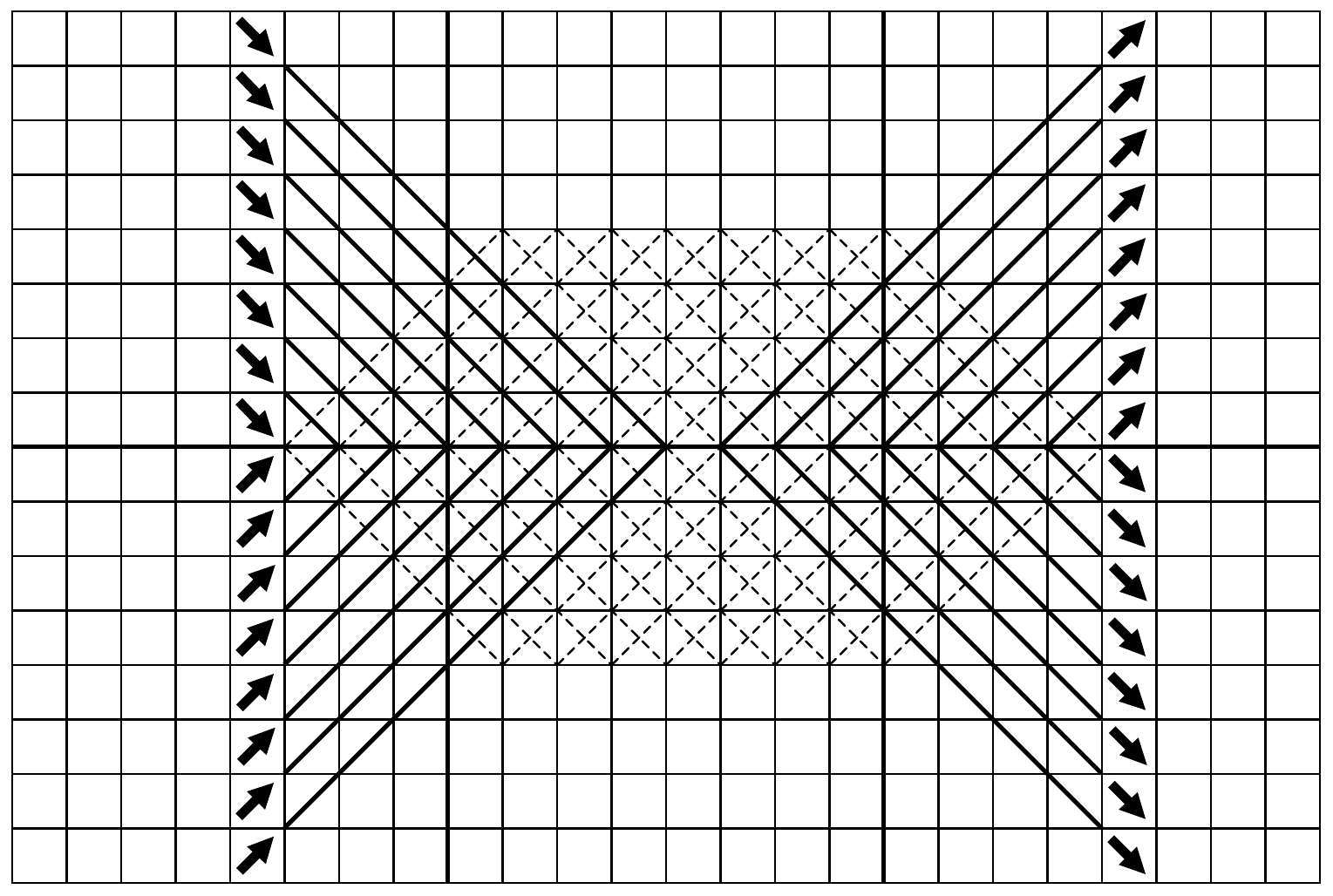}}
{A larger collision in RSS dynamics.  Two columns of 8 particles
  collide.  Partitions are centered at middles and corners of the
  squares shown.}

\subsection{Square balls}

Figure~\ref{fig.rel-block} shows a continuous time history for a
collision of two groups of 8 incoming particles (black arrows on the
left).  The solid lines trace the paths of black particles, the dotted
lines show the paths of white particles.  Interactions occur at both
the centers and the corners of the squares.  The grid is shown with
8$\times$8 blocks of cells outlined by darker lines.  Notice that the
upper group of 8 incoming black particles occupies the fifth column
within an 8$\times$8 block, and the corresponding outgoing group of
black particles also occupies the fifth column within an 8$\times$8
block.  Because of the uniform horizontal motion in this diagram, if
we filled all of the columns in the two 8$\times$8 blocks on the left
of the diagram with black particles (moving right and down in the top
block, right and up in the bottom block), each input column would
separately turn into the corresponding output column within the
rightmost 8 columns.  Thus two 8$\times$8 ``square balls'' would
collide, reconstituting themselves one ball-width to the right of
where the balls would have gone if no collision had occurred.  Thus
the RSS dynamics implements the Soft Sphere collision using square
balls.  By making balls square we allow them to sometimes collide
along a horizontal axis and sometimes along a vertical axis---this
couples the two dimensions.

\subsection{Rescaling}\label{sec.rescaling}

If the square balls were larger, they would shift correspondingly
further to the right: the block dynamics is scale invariant.  The
square ball dynamics scales all the way down to 1$\times$1 square
balls: even single particles reproduce SSM collisions.  We can take a
Soft Sphere Model dynamics that is simulated at the finest grain size
of the dynamics and reproduce it in a system $m$ times larger in each
dimension at a rate $m$ times slower by simply replacing each block of
the partition in the fine grain initial state by a tesselation of
$m\times m$ identical blocks in the scaled system (cf. rescaling
discussion in \cite{marg-phys-like}).

Rescaling can be used as a way to take the continuum limit of an RSS
model.  If the time scale of the underlying RSS dynamics is $m$ times
smaller than that of a scaled system, for $m$ sufficiently large the
microscopic space and time scales can be considered infinitesimal
compared to the ``macroscopic'' scales.  At that point we have the
macroscopic system performing a dynamics that is exactly equivalent to
a microscopic dynamics, but in an effectively continuous space and
time.

When macroscopic blocks collide, there is a question of which block is
which afterwards.  This of course depends on how we decide to
interpret which particle is which when two particles collide.  For
example, if we decide that particle labels never cross the plane of a
two-particle collision, then block collisions look like
Figure~\ref{fig.ssm} (repulsive collision).  If we decide particle
labels always cross, then entire blocks pass through each other as
they come together and then again as they come apart (attractive
collision).

\myfig{rel-mass}{%
\begin{array}{c@{\hspace{.6in}}c@{\hspace{.6in}}c}
\includegraphics[height=1.25in,width=2in]{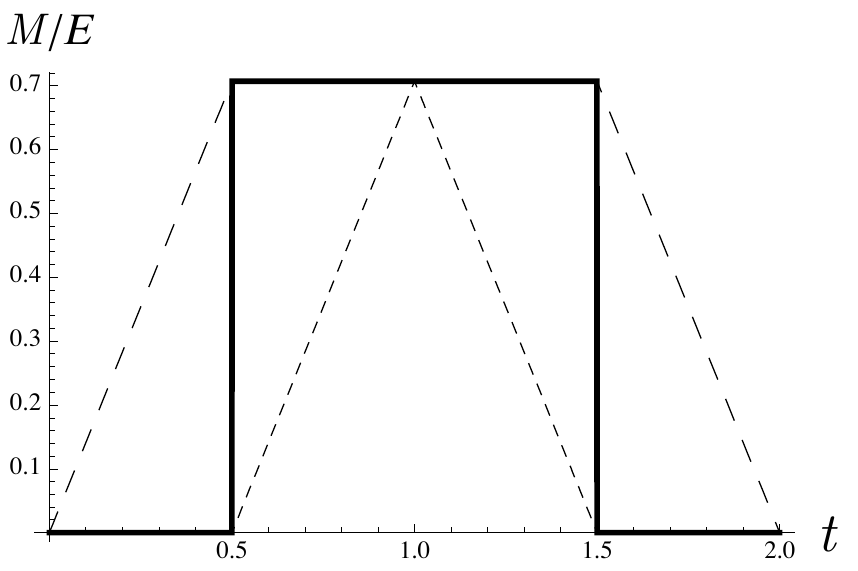} &
\includegraphics[height=1.25in,width=2in]{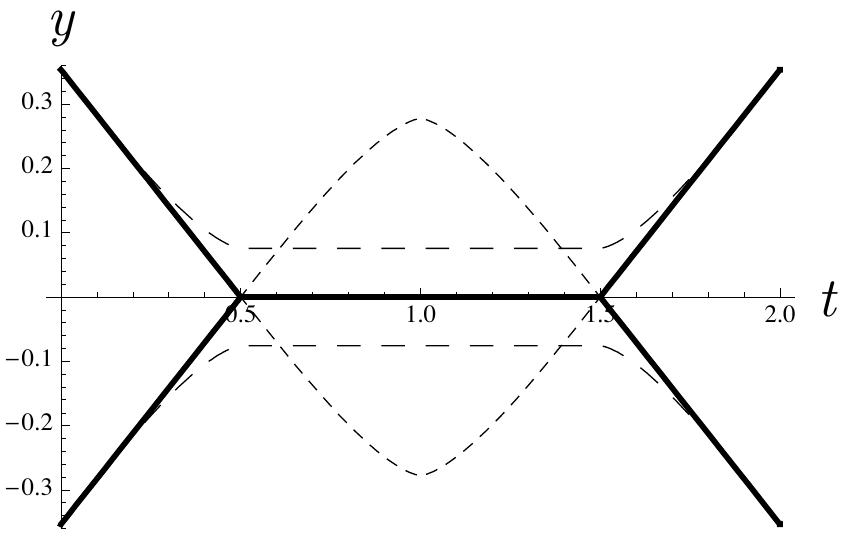} &
\includegraphics[height=1.25in]{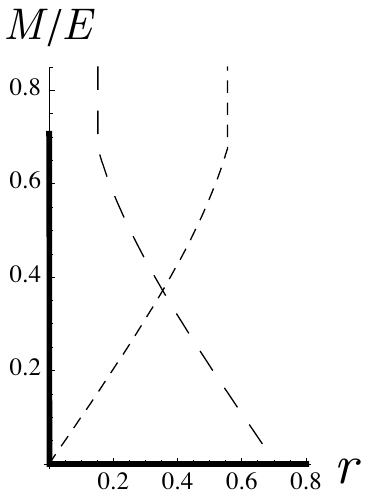} \\ 
\mbox{\bf (a)} & \mbox{\bf (b)} & \mbox{\bf (c)}  \\
\end{array}}
{Macroscopic limit of RSS collision (relativistic).  (a) Fraction of
  energy that is mass, as a function of time.  Three different
  conventions defining mass energy correspond to three different
  time-histories (solid-line, fine-dashed, coarse-dashed).  (b)
  Vertical position $y$ of centers of the top and bottom balls as a
  function of time.  (c) Fraction of energy that is mass, as a
  function of separation of ball centers.}

\subsection{1D and 3D}

In Figure~\ref{fig.rel-block}, we analyzed the collision of 2D blocks
by observing that each column collides independently, as a 1D system.
We could of course simply reinterpret the horizontal axis in the
diagram as time, and this becomes a diagram of a 1D collision of two
extended objects.

In this case blocks in the transition rule of
Figure~\ref{fig.rel-coll}b should be half as wide: particles have only
two directions (up and down).  This is the only ``rotation'' of the
rule in 1D.  We still need two white particle types so that the white
particles can pass through other white particles of the same type
until the two different types come back together and collide at the
``axis'' of the collision.

To get the same rescalable SSM behavior in a 3D system we can simply
apply the planar rule of Figure~\ref{fig.rel-coll}b in various
orientations: only pairs of particles collide and the entire collision
process takes place in the plane defined by the two particle
directions.  We need to use more kinds of labels for the white
particles (more possibilities of which way they can turn), but for
each collision plane only two kinds of turners are needed---these
become the $+$ and $-$ particles in the rule.  Of course the scalable
2D square balls become 3D cubical balls, and we only use a set of
directions for which collisions occur between parallel faces of cubes.

\subsection{Mass energy}\label{sec.mass-energy}

Since the RSS model is a single-speed dynamics that conserves energy
and momentum in its collisions, we are free to interpret it as a
special case of a continuous classical relativistic dynamics with
constrained initial conditions.

As two particles cross over the same spot in
Figure~\ref{fig.rel-block}, we can consider the continuation of each
particle to be either the one that went straight or the one that
turned.  In the former case the particles move at the speed of light
and are massless.  In the latter case the net motion of each particle
during the timestep is horizontal at less than the speed of light and
so it has mass.

In this model all isolated freely moving particles have the same
energy $\epsilon$, whose minimum possible value (Equation~\ref{eq.bce})
sets the ideal energy scale.  For a classical particle moving at the
speed of light, $p=\epsilon/c$.  In this section we use units in which
$c=1$, so that the magnitude of the momentum equals the energy for a
massless particle.  For each particle that changes direction in a
crossover the average momentum during the timestep is only
$\epsilon/\sqrt{2}$ and so the mass is also $\epsilon/\sqrt{2}$, since
$m^2=\epsilon^2-p^2$.

Thus during the collision of Figure~\ref{fig.rel-block}, the mass goes
from zero to a maximum of $8\sqrt{2}\epsilon$ and then back to zero.
In units of $\sqrt{2}\epsilon$, the mass at each step of the collision
is just the number of mass-generating crossings in a column.  A
collision of two $n\times n$ square balls corresponds to $n$
simultaneous collisions such as the one in the figure.  As we let the
balls get larger and larger, the fraction of the total energy that is
mass energy approaches one of the three curves shown in
Figure~\ref{fig.rel-mass}a.

Here we've chosen our unit of time to be the time it takes a freely
moving square ball to move the length of its diagonal.  With this
unit, a collision takes two units of time.  Because of the symmetry of
the collision, the fraction of the energy that is mass is the same for
each of the two colliding balls: each ball changes energy into mass
and then back again according to the curves shown.

Which curve the mass of each ball follows depends on how we interpret
the particle crossings.  We've shown three possible interpretations.
The ``maximum mass'' interpretation, (coarse-dashed line) is given as
a function of time by considering {\em all} crossings to be massive.
The ``sudden mass'' interpretation (thick-solid line) interprets all
particles of each square ball as moving straight (with perhaps a
change in particle-type) until the moment the two balls completely
overlap, and then all crossings have mass until the two balls start to
separate.  The ``white mass'' interpretation (fine-dashed line) is
given by only counting pairs of white particles that cross to be
massive.  The ``maximum mass'' interpretation corresponds to the
minimum possible motion (least kinetic energy).

Let us associate the entire mass of each ball with a single
representative point.  For a freely moving ball, the representative
point is the center of the ball.  Since $E=\gamma M$ the ratio of mass
to energy shown in Figure~\ref{fig.rel-mass}a is equal to $1/\gamma$.
Knowing $\gamma$ as a function of time gives us the velocity $v$ of
the ball as a function of time.  Since the horizontal component of $v$
is unaffected by the collision and remains constant this gives us the
vertical velocity as a function of time, and hence the vertical
position as a function of time.  This is depicted in
Figure~\ref{fig.rel-mass}b, where we've shown the path that the
representative point for each ball must follow during a collision if
its mass varies according to one of the three curves in
Figure~\ref{fig.rel-mass}a.  The three cases show that exactly the
same collision can be interpreted as attractive, repulsive, or sticky.

It is clear from the behavior of $M/E$ and $y$ as functions of time
that the ratio $M/E$ depends only on the separation $r$ between the
two representative points (once the interaction starts).  This
dependence is graphed in Figure~\ref{fig.rel-mass}c.  Thus if we think
of mass as a kind of relativistic potential energy (i.e., total energy
minus kinetic energy), we see that it depends only on the distance
between the two balls.  The potential is zero when the balls are not
yet close enough to touch and it increases as the balls collide.
The maximum mass $M$ for each ball is the fraction $E/\sqrt{2}$ of the
ball's energy, which obtains when all particles are paired so that
there is maximum cancellation of vertical momentum components.

\subsection{Relativistic invariance?}

We were able to use the RSS model to discuss mass in a relativistic
collision by interpreting the model's single particle-speed as the
speed of light and its conservations as relativistic conservations.
This model is, however, clearly not very relativistic even in a
continuum limit.  While it does display macroscopic mechanical
behavior, it does so only in a single inertial frame.  It thus does
not exhibit relativistic frame invariance.

We might expect that, for a colliding pair of RSS particles, the
relativistic kinetic energy (which includes only energy related to net
translational motion) should be closely related to the state-change
energy (which is the minimum energy for the net motion).  This
suggests that we might be able to use the intrinsic definition of
ideal kinetic energy provided by state-change to expose the
non-relativistically invariant character of the RSS model.

If we interpret freely moving RSS particles as massless, as we did in
Section~\ref{sec.mass-energy}, then state-change energy and
relativistic kinetic energy evolve differently.  The state-change
energy of both a free particle and of a block where two particles
cross are the same (Equation~\ref{eq.bce}).  In constrast, if the
relativistic kinetic energy of a free massless particle is $\epsilon$,
then the relativistic kinetic energy of a block containing two
massless particles whose paths intersect at right angles is only
$2\epsilon(1-1/\gamma)\approx .6\epsilon$.

We can try to reinterpret the RSS dynamics to make the relativistic
kinetic energy match the state-change energy both for a free particle
and for a two-particle collision.  If we let free RSS particles each
have mass $m_f$ and let a two-particle collision have mass $m_c$ we
find that the only choice of masses that allows the two energies to be
equal is $m_c=2m_f$, which corresponds to $\gamma=1$.  Thus the
state-change energy becomes the ideal kinetic energy in the RSS model
only for a non-relativistic interpretation of the dynamics.

\section{Conclusions}

A finite-sized QM system has only a finite number of possible distinct
states and can move between distinct states at only a finite rate.
This means that classical special cases of QM dynamics must be
classical finite-state dynamics.  Conversely, we can regard the
dynamics of real physical systems to be a generalization of classical
finite-state dynamics.  This is not a common viewpoint among
physicists today.

The study of classical finite-state dynamics that are special cases of
QM dynamics should, at the least, be of interest for pedagogical
reasons since it allows physical concepts to be seen in an intuitive
classical setting.  This study should also be of interest to
theoretical computer science since it establishes an exact
correspondence between a classical computation and an ideal quantum
realization.  Finally, this study should be of interest to theoretical
physics since it provides a novel finite-state perspective on the
foundations of both classical and quantum mechanics, along with an
intuitive starting point for the construction of new physical models.

\section*{Acknowledgments}

I'd like to thank Tom Knight, Gerry Sussman, Jeff Yepez and Charles
Bennett for their interest and for helpful discussions and comments.

\end{document}